\documentclass[11pt]{article}
\pdfoutput=1

\usepackage{a4wide}
\usepackage[fleqn]{amsmath}
\usepackage{hyperref}

\RequirePackage{amsmath,amssymb,amsxtra,amsthm}

\RequirePackage[T1]{fontenc}
\RequirePackage[utf8]{inputenc}

\RequirePackage{mathrsfs}
\RequirePackage{mathpazo}

\RequirePackage{setspace}
\RequirePackage{array}
\RequirePackage{booktabs}

\RequirePackage{braket}

\RequirePackage[pdftex,final]{graphicx}
\graphicspath{{Plots/}}

\RequirePackage{colortbl}
\RequirePackage{xcolor}
\colorlet{titlerowcolor}{gray!15}
\newcommand*{\tcolrow}{\rowcolor{titlerowcolor}}

\usepackage{color}
\newcommand*{\mathcolor}{}
\def\mathcolor#1#{\mathcoloraux{#1}}
\newcommand*{\mathcoloraux}[3]{%
  \protect\leavevmode
  \begingroup
    \color#1{#2}#3%
  \endgroup
}

\usepackage{cite}

\newcommand{\be}{\begin{equation}}
\newcommand{\ee}{\end{equation}}
\newcommand{\bea}{\begin{eqnarray}}
\newcommand{\eea}{\end{eqnarray}}

\newcommand{\cF}{\mathcal{F}}

\newcommand{\cH}{\mathcal{H}}

\newcommand{\cN}{\mathcal{N}}

\newcommand{\cZ}{\mathcal{Z}}

\numberwithin{equation}{section}
\numberwithin{table}{section}
\numberwithin{figure}{section}

\author{
  \begin{minipage}{0.97\linewidth}
    \vspace{1cm}
    \begin{center}
      \begin{small}
        \textbf{Carlo Angelantonj$^a$, Ioannis Florakis$^b$, Mirian Tsulaia$^c$}
     \end{small}
    \end{center}
    \vspace{.3cm} \hspace{1.3cm}\begin{minipage}{.75\linewidth}
     {\it \begin{footnotesize}
          \begin{itemize}
          \item[${}^a$] Dipartimento di Fisica, Universit\`a di Torino, and INFN Sezione di Torino
          \\
            Via P. Giuria 1, 10125 Torino, Italy
          \item[${}^b$] CERN Theory Unit, 1211 Geneva 23, Switzerland
          \item[${}^c$] Faculty of Education Science Technology and Mathematics, \\
          University of Canberra, Bruce ACT 2617, Australia
          \end{itemize}
        \end{footnotesize}}
         \end{minipage}
    \vspace{1cm}
  \end{minipage}
}

\date{}

\title{\vspace{3cm}
  \begin{huge}
    \textbf{Generalised universality of gauge thresholds in heterotic vacua with and without supersymmetry}
  \end{huge}
}

\begin{document}

\begin{titlepage}
  \maketitle
  \thispagestyle{empty}

  \vspace{-14cm}
  \begin{flushright}
   CERN-PH-TH/2015-211
   \end{flushright}

  \vspace{11cm}

  \begin{center}
    \textsc{Abstract}\\
  \end{center}

We study one-loop quantum corrections to gauge couplings in heterotic vacua with spontaneous supersymmetry breaking. Although in non-supersymmetric constructions these corrections are not protected and are typically model dependent, we show how a universal behaviour of threshold differences, typical of supersymmetric vacua, may still persist. We  formulate specific conditions on the way supersymmetry should be broken for this to occur. Our analysis implies a generalised notion of threshold universality even in the case of unbroken supersymmetry, whenever extra charged massless states appear at enhancement points in the bulk of moduli space. Several examples with universality, including non supersymmetric chiral models in four dimensions, are presented.

\vfill

{\small
\begin{itemize}
\item[E-mail:] {\tt carlo.angelantonj@unito.it}\\ {\tt ioannis.florakis@cern.ch}\\
{\tt mirian.tsulaia@canberra.edu.au}
\end{itemize}
}

\end{titlepage}

\setstretch{1.1}

\tableofcontents


\vskip 1in


\section{Introduction}

The way to break supersymmetry spontaneously in heterotic string theory which admits a solvable Conformal Field Theory description is via the so-called stringy Scherk-Schwarz mechanism \cite{Rohm:1983aq,SSstringii,FKPZ,Kounnas:1989dk}, corresponding to a particular flat gauging of $\cN =4$ supergravity. It amounts to shifting the masses of the perturbative states by their $R$-symmetry charges. It is well known that this deformation can be conveniently reformulated at the world-sheet level in terms of freely-acting orbifolds, where a supersymmetry breaking generator $g_{\rm SB}$ is coupled to a shift along one (or more) of the compact cycles. 

Non-supersymmetric constructions have been extensively studied in the past in heterotic and type II  \cite{AlvarezGaume:1986jb, Ginsparg:1986wr, Nair:1986zn,Dixon:1986iz,SSstringii,FKPZ,Kounnas:1989dk, Itoyama:1986ei,Taylor:1987uv, Toon:1990ij, Sasada:1995wq, Kiritsis:1997ca, Florakis:2009sm, Florakis:2010ty, Faraggi:2007tj, Faraggi:2009xy, Ghilencea:2001bv,Dienes:1995bx, Font:2002pq, Harvey:1998rc}, as well as type I \cite{Antoniadis:1998ki, Antoniadis:1998ep, Antoniadis:1999ux, Angelantonj:1999gm, Angelantonj:2003hr, Angelantonj:2006ut, Angelantonj:2005hs, Borunda:2002ra, AS, Angelantonj:1998gj, Blumenhagen:1999ns, Sugimoto:1999tx, Antoniadis:1999xk, Angelantonj:1999ms, Angelantonj:1999jh, Aldazabal:1999jr, GatoRivera:2007yi, GatoRivera:2008zn} strings and have become the subject of recent investigation in the context of non-supersymmetric string phenomenology \cite{Abel:2015oxa, Blaszczyk:2014qoa, Blaszczyk:2015zta, Nibbelink:2015vha, Ashfaque:2015vta, Lukas:2015kca}.  Criteria for classical stability of non-supersymmetric vacua have been analysed in \cite{Kutasov:1990sv, Dienes:1994np, Angelantonj:2010ic} and reflect the presence of a {\em misaligned supersymmetry} in the spectrum of string excitations whose distribution is governed by the location of the non-trivial zeroes of the Riemann zeta-function in the complex plane. The construction of (lower-dimensional) non-supersymmetric string vacua that are classically stable throughout the entire moduli space is a challenging problem due to the emergence of tachyonic instabilities in certain regions of the classical moduli space, in a way that is reminiscent of the celebrated Hagedorn instability of string thermodynamics \cite{Hagedorn:1965st, Atick:1988si, Antoniadis:1991kh}. Recent studies of orbifold and Calabi-Yau compactifications have shown, however, that some of these pathologies might be lifted when the orbifold singularities are blown up \cite{Blaszczyk:2014qoa}. Furthermore, although supersymmetry breaking typically induces sizeable contributions to the one-loop vacuum energy, models with massless Fermi-Bose degeneracy have been constructed \cite{Harvey:1998rc, Angelantonj:1999gm, Shiu:1998he, Abel:2015oxa}. These only yield exponentially suppressed contributions in the volume of the compact space, thus softening the back-reaction to the classical vacuum. Moreover, progress has also been made in the attempt to construct non-supersymmetric heterotic vacua with semi-realistic spectra.

Although, these analyses have been primarily conducted in a case-by-case fashion, in this paper we are rather interested in disclosing some model-independent features of heterotic vacuum configurations with spontaneously broken supersymmetry. In particular, we shall focus on the study of one-loop corrections to gauge couplings in the low-energy effective action, generalising the analysis of \cite{AFT}. 

Indeed, in \cite{AFT} we uncovered a remarkable universal structure in the moduli dependence of differences of threshold corrections $\varDelta_{\alpha\beta}$ in heterotic orbifold compactifications where supersymmetry is spontaneously broken by a four-dimensional analogue of the Itoyama-Taylor deformation \cite{Itoyama:1986ei}. This extends and generalises the familiar universality of supersymmetric constructions \cite{Kiritsis:1996dn, Dixon:1990pc, Mayr:1993mq}.

In particular, we found \cite{AFT} that, in a large class of four-dimensional orbifolds with factorisable six-torus, threshold differences assume the {\em universal} expression
\begin{equation}
\begin{split}
\varDelta_{\alpha\beta} &= \sum_{i=1,2,3} \Bigl\{ a_i \, \log \, \left[ T_2^{(i)} U_2^{(i)} \, |\eta (T^{(i)}) \eta (U^{(i)}) |^4 \right] 
\\
&\qquad + b_i \, \log \, \left[ T_2^{(i)} U_2^{(i)} \, |\vartheta_4 (T^{(i)}) \vartheta_2 (U^{(i)}) |^4 \right] 
+ c_i \, \log |j_2 (T^{(i)}/2 ) - j_2 (U^{(i)}) |^4 \Bigr\}\,,
\end{split}
\label{universal}
\end{equation}
where $T^{(i)}$ and $U^{(i)}$ are the K\"ahler and complex structure moduli of the $i$-th two-torus in the $T^6 = T^2 \times T^2 \times T^2$ decomposition, $\eta$ is the Dedekind function, $\vartheta_2$ and $\vartheta_4$ are the Jacobi theta constants, and $j_2$ is the analogue of the Klein invariant $j$ function for the $\varGamma_0 (2)$ subgroup of ${\rm SL} (2;\mathbb{Z})$. The only model-dependent quantities are the constants $a_i$, $b_i$ and $c_i$, which can be easily computed from the tree-level massless spectrum. As we shall show, they are related to differences of suitable $\beta$-function coefficients. 

Normally, a generic one-loop string amplitude receives contributions from the whole tower of Kaluza-Klein, winding and oscillator excitations of the closed string. However, the presence of  the universal behaviour \eqref{universal} requires that only oscillators from a chiral sector of the world-sheet theory do contribute, so that holomorphy and modular invariance can drastically constrain the one-loop integrand. Although, this is naturally the case when supersymmetry is present, since the $F_{\mu\nu} F^{\mu\nu}$ coupling is BPS protected, it is far from trivial that this should still hold when supersymmetry is absent. This requirement can be translated into the presence of a spectral flow \cite{Faraggi:2011aw} within the gauge degrees of freedom of the heterotic string, which projects onto the bosonic (right-moving) ground state. This mirrors the property of BPS saturated amplitudes of projecting  onto the Ramond (left-moving) ground state.

Whether  this spectral flow is present or not, clearly depends on the choice of the supersymmetry breaking generator $g_{\rm SB}$. As we shall show, a necessary condition for universality is the existence of a related supersymmetric heterotic vacuum where $g_{\rm SB}$ is actually replaced by a supersymmetry preserving generator with the same action on the gauge degrees of freedom. Moreover, the supersymmetric vacuum ought not give rise to extra massless states at special points in the bulk of the $(T,U)$ plane. These conditions strongly constrain the possible ways of breaking supersymmetry. 

The scope of the  present investigation is threefold. Firstly, it is to explain the origin of this universal behaviour, which is unexpected when supersymmetry is absent. Secondly, it is to clearly define the explicit conditions for this to occur. Thirdly, it is to construct further examples exhibiting this universal structure. These include four-dimensional chiral models which open the possibility for a more phenomenologically oriented investigation.

The paper is organised as follows: in Section 2 we present general properties of orbifold constructions, introduce the decomposition of the associated partition functions into modular orbits. Section 3 contains our main result on the universality structure of threshold differences for generic heterotic vacua with and without supersymmetry and provides the precise conditions for it to occur. Section 4 contains explicit examples of thresholds for vacua with spontaneously broken supersymmetry, including a four-dimensional chiral vacuum with $\cN=1 \to \cN=0$. 
In Sections 5 and 6 we elaborate on ways of departing from universality depending on which conditions of the universality theorem are violated. In Section 7 we provide a brief discussion on the scales of supersymmetry breaking, while conclusions are given in Section 8.

\section{Dissecting an orbifold}\label{dissect}

\begin{figure}
\centerline{
\includegraphics[width=12cm]{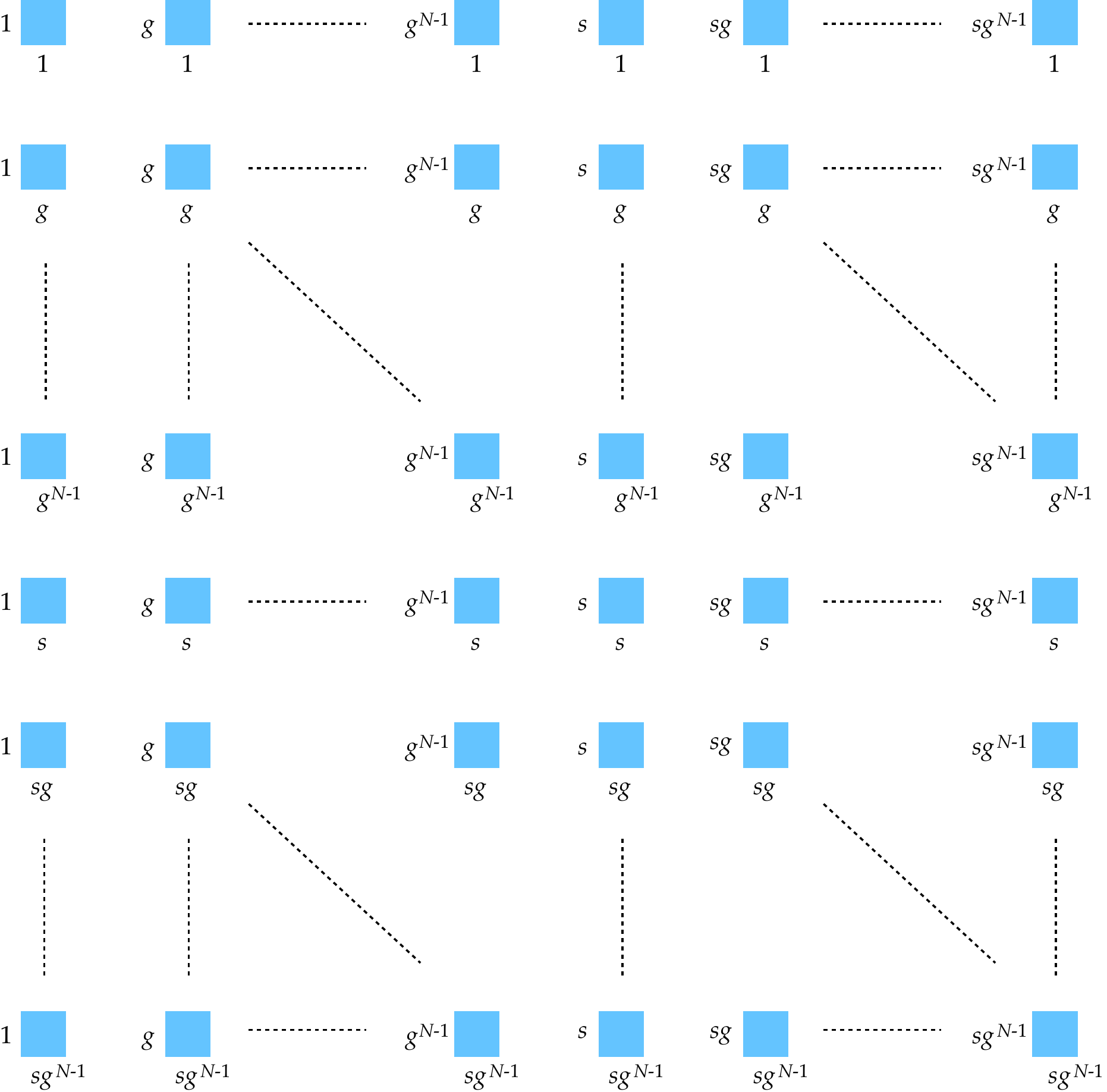}
}
\caption{The orbit structure of a generic $\mathbb{Z}_N \times \mathbb{Z}_2$ orbifold. Boxes schematically represent the world-sheet torus with the proper length $\sigma$ running along the horizontal side and the proper time $\tau$ running along the vertical side. The horizontal label $\alpha$ denotes the $\alpha$-twisted sector of the strings, while the horizontal label $\beta$ denotes the insertion of the $\beta$ element in the trace.}\label{boxes}
\end{figure}

\begin{figure}
\centerline{
\includegraphics[width=6cm]{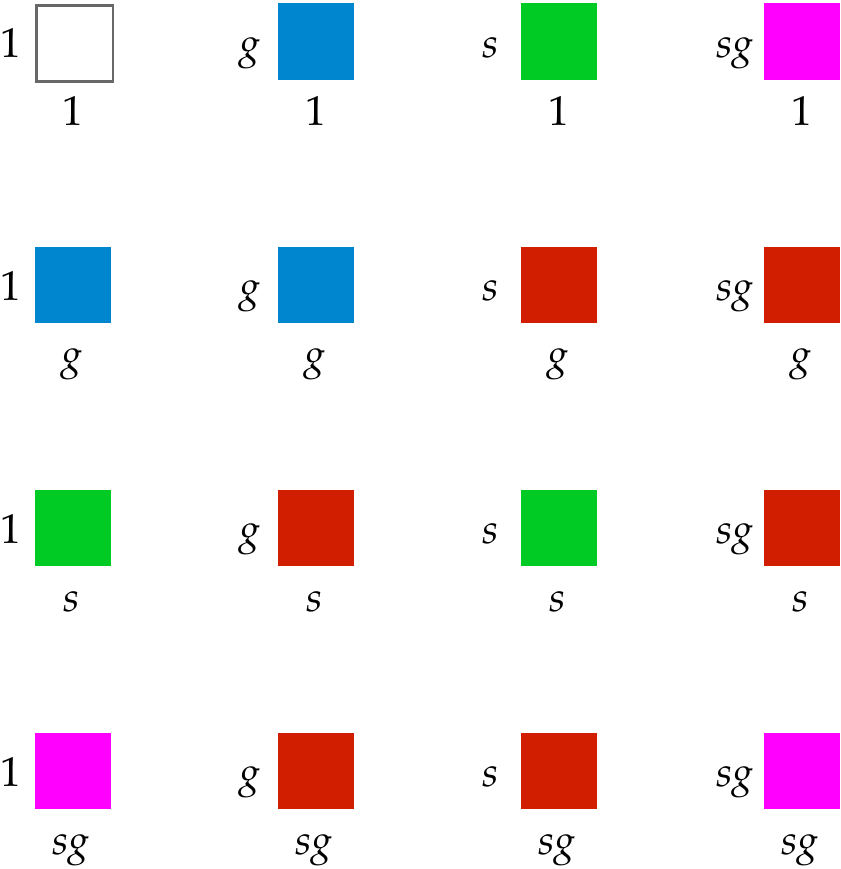}
}
\caption{The orbit structure of the $\mathbb{Z}_2 \times \mathbb{Z}_2$ orbifold. Boxes are coloured according to the modular orbit they belong to.}\label{boxes2}
\end{figure}

In order to unveil the universality property of threshold corrections in heterotic vacua with spontaneously broken supersymmetry, it is instructive to dissect orbifold compactifications that actually preserve $\cN =2$ supersymmetry. The advantage of this pursuit will become transparent in the next section. To be specific, we shall restrict our attention to the heterotic ${\rm E}_8 \times {\rm E}_8$ string compactified on the $T^6 / \mathbb{Z}_N \times \mathbb{Z}_2$ orbifolds, where the $\mathbb{Z}_N$ is generated by the element $g$ and realises the singular limit of K3, whereas the $\mathbb{Z}_2$, generated by the element $s$, is freely acting and preserves the original $\cN =4$ supersymmetries, {\em i.e.}  only a non-trivial action on the gauge sector, corresponding to discrete Wilson lines, is allowed.
It is convenient to represent pictorially these orbifold compactifications as in figure \ref{boxes}. Each box in the figure with horizontal label $\alpha$ and vertical label $\beta$ represents the orbifold block
\begin{equation}
\cZ \big[{\textstyle{\alpha \atop \beta}}\big] = {\rm Tr}_{\cH_\alpha} \beta \, q^{L_0-c/24}\, \bar q^{\bar L_0 -\bar c/24}\,,
\end{equation}
where $\alpha$ and $\beta$ stand for generic elements of the $ \mathbb{Z}_N \times \mathbb{Z}_2$ orbifold, and $\cH_\alpha$ denotes the Hilbert space in the sector twisted by $\alpha$. Although the full partition function is invariant under the modular group ${\rm SL} (2;\mathbb{Z})$, each individual box is only invariant with respect to some finite index subgroup. Moreover, the various boxes are conveniently arranged in orbits of the modular group. For concreteness, we now focus on the $N=2$ case, depicted in figure \ref{boxes2}. Here we have five independent modular orbits. The white one is trivial and corresponds to the toroidal compactification of the heterotic string. The blue one corresponds to the non-trivial K3 subsectors and is generated by action of $1, S, TS$ transformations on the single element $\cZ [{1\atop g}]$ which is invariant under the Hecke congruence subgroup $\varGamma_0 (2)$ \cite{AFP3}. Similarly for the green and magenta modular orbits which are generated by the elements $\cZ [{1\atop s}]$ and $\cZ [{1\atop sg}]$, respectively. The red orbit is special because it cannot be obtained by the action of ${\rm SL} (2;\mathbb{Z})/\varGamma_0 (2)$ on a single element. Instead it is generated by the action of $1, S$ and $TS$ on the sum $\cZ [{g\atop s}] + \cZ [{g\atop sg}]$. Similar orbit decompositions occur also for the other K3 realisations, though in those cases more orbits, associated to different Hecke congruence subgroups, are present \cite{AFP3}.

The partition function encodes not only the physical (level-matched) spectrum of the theory, but modularity implies that unphysical (non-level-matched) states also be counted. Among the unphysical states particularly important for our investigation are those associated to the vacuum of the right moving sector. These include the ubiquitous untwisted uncharged bosonic vacuum $|0\rangle$ of the heterotic string, as well as the vacua 
\begin{equation}
|n \rangle \sim \bar \sigma_n \, \bar \chi_n \,  |0\rangle\,, \qquad n = 1\,, \ldots \,, N-1\,,
\end{equation}
associated to the right-moving ground states of the $g^n$ twisted sectors. $\sigma_n$ is the standard bosonic twist field associated to the right-moving twisted K3 coordinates and has conformal weight $\varDelta_\sigma=\frac{n}{N}(1-\frac{n}{N})$, whereas $\chi_n$ is the fermionic twist field associated to the ``twisted Kac-Moody current'' and has ${\rm U} (1)$ charge $Q_\chi = \pm \frac{n}{N}$ and conformal weight $\varDelta_\chi = Q_\chi^2$ \cite{Dixon:1986qv,Hamidi:1986vh}. We stress once more that $|0\rangle$ is neutral with respect to the ${\rm E}_8 \times{\rm E}_8$ Cartan charges, while $|n\rangle$ always carries non-trivial charge with respect to one of these Cartan's, at least in the case of standard embedding. 

Although in most discussions one discards such unphysical states, unitarity demands that they do contribute to loop amplitudes and, as we shall soon see, they play an important r{\^o}le in controlling the universality structure of threshold differences in heterotic vacua with spontaneously broken supersymmetry.

\section{The anatomy of gauge threshold corrections}

Threshold corrections in heterotic string vacua have been the subject of extensive study in the nineties \cite{Dixon:1990pc, Kiritsis:1996dn, Mayr:1993mq, Kaplunovsky:1987rp, Mayr:1995rx, Antoniadis:1992pm, Gregori:1997hi, Harvey:1995fq,LopesCardoso:1996nc,LopesCardoso:1994vn}. For a gauge group factor $G_\alpha$ with Kac-Moody level $k_\alpha$, the running of the effective coupling $g_\alpha$ at a scale $\varLambda$ is dictated by 
\begin{equation}
\frac{16\pi^2}{g_\alpha^2 (\varLambda )} = \frac{16\pi^2\, k_\alpha}{g_s^2} + \beta_\alpha \, \log \frac{M_s^2}{\varLambda^2} + \varDelta_\alpha\,,
\end{equation}
where the logarithmic term accounts for the massless degrees of freedom with $\beta$-function coefficient $\beta_\alpha$, whereas the contribution of the infinite tower of massive string modes is encoded in the threshold correction $\varDelta_\alpha$. Here $g_s$ is the string coupling constant and $M_s$ is the string scale.

Regardless of the presence or not of space-time supersymmetry, the one-loop heterotic thresholds $\varDelta_\alpha$ are given by \cite{Kaplunovsky:1987rp}
\begin{equation}
\varDelta_\alpha = {\rm R.N.}\, \int_\cF d\mu \, \frac{i\tau_2}{\pi\, \eta^2 \bar\eta^2} \sum_{a,b} \partial_\tau \left( \frac{\theta \bigl[{\textstyle{a/2\atop b/2}}\bigr]}{\eta}\right) \, {\rm Tr}_{\cH{[{a\atop b}]}} \, \left[\left( Q_\alpha^2 - \frac{k_\alpha}{4\pi\tau_2}\right)\, q^{L_0-c/24}\, \bar q^{\bar L_0 -\bar c /24}\right] \,, \label{thresholds}
\end{equation}
where $q=e^{2i\pi\tau}$, $\theta [{a\atop b}]$ are Jacobi theta constants with characteristics, and the trace runs over the Hilbert space $\cH \big[{\textstyle{a\atop b}}\big]$ of the internal $(c,\bar c) = (9,22)$ CFT system with given spin structures. The trace is weighted with a Cartan charge $Q_\alpha$, while the presence of  $\tau_2^{-1}$  is ascribed to a contact term in the correlation function of the Kac-Moody currents. It is independent of the gauge group charges and can be associated to the universal coupling of the dilaton. The sum is restricted to the even spin structures $(a,b) \not= (1 , 1)$. The modular integral with measure $d\mu = \tau_2^{-2} d\tau_1 d\tau_2$ is to be performed over the ${\rm SL} (2;\mathbb{Z})$ fundamental domain $\cF$, and we invoke the modular-invariant regularisation prescription of \cite{AFP1,AFP2,AFP3} to treat the infra-red divergences ascribed to the massless string states. Henceforth, we shall not explicitly display the ${\rm R.N.}$ symbol in front of modular integrals, but we shall tacitly assume that all integrals be regularised according to \cite{AFP1,AFP2,AFP3}. 

When computing the difference of thresholds $\varDelta_{\alpha\beta} = \varDelta_\alpha - \varDelta_\beta$, the $\tau_2^{-1}$ contact term cancels out, and in generic orbifold compactifications without continuous Wilson lines, they take the schematic form
\begin{equation}
\varDelta_{\alpha\beta} = \int_\cF d\mu \, \sum_{h,g} \, L \big[{\textstyle{h\atop g}}\big] (\tau ) \, \bar\varPhi \big[{\textstyle{h\atop g}}\big] (\bar\tau )\, \varGamma \big[{\textstyle{h\atop g}}\big] (G,B) \,, \label{standardthre}
\end{equation}
where the sum runs over the various sectors of the orbifold. $L \big[{\textstyle{h\atop g}}\big] (\tau ) $ is an holomorphic function of the modulus $\tau$ encoding the helicity super-trace over the left-moving sector, $\bar\varPhi \big[{\textstyle{h\atop g}}\big] (\bar\tau )$ is an anti-holomorphic function encoding the $Q^2_{\alpha\beta} = Q^2_\alpha - Q^2_\beta$ graded trace over the right-moving sector, while $\varGamma \big[{\textstyle{h\atop g}}\big] (G,B)$ denotes a generic Narain lattice partition function associated to shifted tori with metric $G$ and $B$-field backgrounds.

As anticipated, in the present work we shall confine our attention to the moduli dependence of $\varDelta_{\alpha\beta}$, since constant contributions are ambiguous and depend on the infra-red renormalisation scheme. In supersymmetric compactifications, the only moduli dependence of gauge thresholds originates from those sectors preserving $\cN=2$ supersymmetry\footnote{The untwisted unprojected sector of any supersymmetric orbifold compactification, although depending on the moduli of the six-dimensional torus, does not contribute to the threshold since it preserves the full $\cN =4$ supersymmetry.}, which correspond to a ${\rm K3}\times T^2$ orbifold.  These sectors have the remarkable property that their holomorphic contribution $L \big[{\textstyle{h\atop g}}\big]$ drops to a constant, as a consequence of the BPS-ness of the $F_{\mu\nu} F^{\mu\nu}$ coupling. Technically, this property is the result of a  cancellation between the helicity super-trace and the holomorphic contribution of the twisted K3 lattice, so that only the left-moving ground state survives. As a result, for a generic $T^4/\mathbb{Z}_N$ realisation of the K3 surface, one has
\begin{equation}
\varDelta_{\alpha\beta} = \int_\cF d\mu \, \sum_{{\textstyle{h,g=0 \atop (h,g)\not= (0,0)}}}^{N-1} \bar\varPhi \big[{\textstyle{h\atop g}}\big] (\bar\tau )\, \varGamma_{2,2} \big[{\textstyle{h\atop g}}\big] (T,U) \,,
\label{Deltagen}
\end{equation}
where we allow for the possibility that the $\cN =4 \to \cN=2$ breaking be spontaneous, and realised via a freely acting orbifold, whereby the $\mathbb{Z}_N$ rotation on the $T^4$ is coupled to a translation along a one-cycle of the  $T^2$. The Narain partition function associated to the shifted lattice then reads
\begin{equation}
\varGamma_{2,2}  \big[{\textstyle{h\atop g}}\big] (T,U) = \tau_2\, \sum_{m_i , n^i \in \mathbb{Z}} e^{\frac{2 i \pi g}{N} (m_i \lambda^i + n^i \mu_i ) }q^{\frac{1}{4} |p_{\rm L}|^2 } \, \bar q ^{\frac{1}{4} |p_{\rm R}|^2}\,,
\end{equation}
with the lattice momenta given by
\begin{equation}
\begin{split}
p_{\rm L} &= \frac{1}{\sqrt{T_2 U_2}} \left[ \hat m_2 - U \hat m_1 + \bar T ( \hat n^1 + U \hat n^2 )\right]\,,
\\
p_{\rm R} &=\frac{1}{\sqrt{T_2 U_2}} \left[ \hat m_2 - U \hat m_1 +  T ( \hat n^1 + U \hat n^2 )\right]\,,
\end{split}
\end{equation}
where $\hat m_i = m_i + \frac{h}{N} \lambda_i$ and $\hat n ^i = n^i + \frac{h}{N} \mu^i$, for generic momentum and winding shifts $\lambda_i , \mu^i \in\mathbb{Z}_N$.

The functions $ \bar\varPhi \big[{\textstyle{h\atop g}}\big] (\bar\tau )$ are highly constrained by modular invariance, which is the rationale behind the celebrated universality of differences of gauge thresholds \cite{Dixon:1990pc,Kiritsis:1996dn}. In fact, whenever the lattice of the $T^2$ factorises, {\em i.e.} when $\lambda_i = 0 = \mu^i$, as in hard $\cN =4 \to \cN=2$ breaking without Wilson lines, it was realised that
\begin{equation}
\bar\varPhi \equiv \sum_{{\textstyle{h,g=0 \atop (h,g)\not= (0,0)}}}^{N-1} \bar\varPhi \big[{\textstyle{h\atop g}}\big] (\bar\tau ) = {\rm const} \,,
\label{naive}
\end{equation}
thus yielding the celebrated result in \cite{Dixon:1990pc}. Indeed, in this case $\varPhi$ is a holomorphic function invariant under the full ${\rm SL} (2;\mathbb{Z})$ modular group which is bound to be regular at the cusp $\tau= i \infty$. This last condition is the consequence of the fact that the untwisted bosonic vacuum $|0\rangle$ of the heterotic string is neutral with respect to the Kac-Moody currents while the charged twisted right-moving vacua $|n\rangle$ are not invariant under the orbifold action. Notice that this does not necessarily imply that each $\bar\varPhi \big[{\textstyle{h\atop g}}\big]$ be a constant for fixed $h$ and $g$. For instance,
$\bar\varPhi \big[{\textstyle{0\atop 1}}\big]$ is only invariant under the Hecke congruence subgroup $\varGamma_0 (N)$, and the space of holomorphic invariant functions which are regular at the cusp $\tau=i\infty$ is much richer. In fact, one has the general decomposition 
\begin{equation}
\varPhi \big[{\textstyle{0\atop 1}}\big] (\tau) = a + \sum_{\mathfrak{a}\not=\infty} b_\mathfrak{a} \, j_\mathfrak{a} (\tau) \,,
\label{Heckedecomp}
\end{equation}
where $a$ and $b_\mathfrak{a}$ are constants and are the only model-dependent data. The index $\mathfrak{a}$ labels the various cusps of the fundamental domain $\cF_N$ of $\varGamma_0 (N)$, and $j_\mathfrak{a}$ are the invariant functions attached to the cusp $\mathfrak{a}$. They are related to the Hauptmodul $j_\infty (\tau) $ via 
\begin{equation}
j_\mathfrak{a} (\tau ) = j_\infty (\sigma_\mathfrak{a}\, \tau )\,.
\end{equation}
The ${\rm SL} (2;\mathbb{R})$ matrices $\sigma_\mathfrak{a}$ are involutions that relate the various cusps of $\varGamma_0 (N)$. Notice that only the Hauptmodul $j_\infty$ has a simple pole $q^{-1}$, whereas all others are regular at the infra-red point. We refer the reader to \cite{AFP3} for more details on the structure of the fundamental domain and modular forms of Hecke congruence subgroups. In the following we shall only need the $\varGamma_0 (2)$ case, which has two cusps at $\tau = i\infty $ and $\tau = 0 $, related by the Fricke involution
\begin{equation}
\sigma_0 = \begin{pmatrix} 0 & 1/\sqrt{2} \\ - \sqrt{2} & 0\end{pmatrix}\,.
\end{equation}
The Hauptmodul and its Fricke transform can be expressed in terms of the familiar elliptic functions as
\begin{equation}
j_\infty (\tau ) \equiv j_2 (\tau ) = \left( \frac{\eta (\tau)}{\eta (2\tau )}\right)^{24}+24\,, \qquad 
j_0 (\tau ) \equiv \hat\jmath_2 = \left( \frac{\vartheta_2 (\tau ) }{\eta (\tau ) }\right)^{12}+24\,.
\end{equation}

The decomposition \eqref{Heckedecomp} dramatically extends the notion of universality in supersymmetric cases, when the two-dimensional lattice is shifted and couples to $\varPhi \big[{\textstyle{0\atop 1}}\big]$. In fact, it is of crucial importance for the universality of related vacua with broken supersymmetry. To appreciate this point, let us focus for simplicity on the  $N=2$ case which corresponds to the $\mathbb{Z}_2$ realisation of K3. Upon partially unfolding $\cF$ \cite{AFP3}, one can cast the integral \eqref{Deltagen} as
\begin{equation}
\begin{split}
\varDelta_{\alpha \beta} &= \int_{\cF_2} d\mu \, \varGamma_{2,2} \big[{\textstyle{0\atop 1}}\big] (T,U) \, \bar\varPhi \big[{\textstyle{0\atop 1}}\big] (\bar\tau )
\\
&= a\,  \int_{\cF_2} d\mu\, \varGamma_{2,2} \big[{\textstyle{0\atop 1}}\big] (T,U) +  b_0\, \int_{\cF_2} d\mu\, \varGamma_{2,2} \big[{\textstyle{0\atop 1}}\big] (T,U)\, \bar \jmath_0 (\bar\tau ) \,.
\end{split}
\end{equation}
The  first integral was evaluated in \cite{Mayr:1993mq,Kiritsis:1998en,AFP3} while the the second one was evaluated in \cite{AFP3,AFP4} and, for a momentum shift along the first cycle of $T^2$, read
\begin{equation}
\varDelta_{\alpha \beta } = - (a+24 \, b_0 )\, \log \left[ T_2 U_2 |\vartheta_4 (T) \vartheta_2 (U) |^4 \right] -2\, b_0 \, \log |j_\infty (T/2 ) - j_\infty (U) |^4\,.
\label{universality}
\end{equation}
Similar expressions can be obtained also for the other orbifold realisations of K3.  While the first contribution is regular in the bulk of the $(T,U)$ moduli space, the second one displays a logarithmic singularity when $T/2 = U$, plus all their $\varGamma_0 (2)_U $ images. Therefore, a non-vanishing $b_0$ coefficient is associated to the presence of extra massless states at these special points. The lesson to be learnt from this discussion, is that the presence of extra charged massless states, whether they be hypermultiplets or vector multiplets, significantly modifies the standard notion of universality  of $\varDelta_{\alpha\beta}$ and extends it with the contribution in eq. \eqref{universality}.

As we shall see momentarily, this very same universal behaviour of supersymmetric gauge thresholds also holds in heterotic models with spontaneously broken supersymmetry \cite{AFT}, provided one considers thresholds for groups of rank larger than one, and certain specific conditions are satisfied by the supersymmetry-breaking generators. To this end, we focus on heterotic vacua on $T^6 / \varOmega_{\rm S} \times \varOmega_{\rm SB}$, where $ \varOmega_{\rm S}$ is a supersymmetric orbifold, while $ \varOmega_{\rm SB}$ has a free action, and is responsible for the spontaneous breaking of supersymmetry. We restrict our analysis to classically stable vacua. Although, it might be difficult to construct non-supersymmetric vacua without tachyonic excitations at any point of the full moduli space \cite{Ginsparg:1986wr,Nair:1986zn}, one can always find regions which are tachyon-free. We shall confine our analysis to these regions, barring questions about quantum stability. Only in the absence of tachyons, can the one-loop perturbative expansion be trusted. In this cases, eq. \eqref{thresholds} still computes the radiative corrections to gauge couplings, and the difference of thresholds is again given by \eqref{standardthre}. 

The main difference with the supersymmetric case, is that now the $F_{\mu\nu} F^{\mu\nu}$ coupling is no longer BPS protected, and receives contributions from the whole tower of string states. This implies that there are sectors where the helicity supertrace no longer cancels against the contribution of the twisted lattice, and therefore the $L [{h\atop g}]$'s are no longer constants for all $h$ and $g$. As a result, aside from lattice contributions, the integrand in \eqref{standardthre} is not holomorphic any more, and modular invariance fails to constrain it uniquely\footnote{Actually, this is not strictly speaking true. Although the integrand is not holomorphic, it is not a general real modular function which indeed is not fully constrained by modular invariance. It is rather a sesquilinear combination of the form $\sum_{ij} C_{ij} \chi_i (\tau )\, \bar \chi_j (\bar \tau) $. Each  $\chi_i$ is  a holomorphic function  invariant under some finite index subgroup of ${\rm SL} (2;\mathbb{Z})$, and is hence strongly constrained. This implies the presence of a much more general notion of universality. However, we shall reserve the term {\em universality} to refer only to the case of holomorphic integrands.}. Universality is recovered whenever the $\varPhi [{h\atop g}]$ are constants for suitable values of $h$ and $g$, 
so that the product $L [{h\atop g}]\, \varPhi [{h\atop g}]$ reduces to a holomorphic or anti-holomorphic function. In this case modular invariance fixes the $\tau$ dependence of the (lattice independent) integrand up to few model dependent coefficients. 

To be concrete, let us give a closer look at the conditions required for universality. We denote by $g_{\rm S}$ and $g_{\rm SB}$ generic elements of the orbifold groups $\varOmega_{\rm S}$ and $\varOmega_{\rm SB}$. Clearly,  $L [{g^\prime_{\rm S}\atop g_{\rm S}}]$ is constant due to the effective supersymmetry present in this orbit, and therefore yields only universal contributions to the thresholds. However, in supersymmetry-breaking orbits $L [{g_{\rm S}\atop g_{\rm SB}}]$ is no longer a constant and thus universality is lost unless $\varPhi  [{g_{\rm S}\atop g_{\rm SB}}]$ is. In this case, modular invariance leads to the decomposition of $L [{g_{\rm S}\atop g_{\rm SB}}]$ into functions invariant under some finite index subgroup of ${\rm SL} (2;\mathbb{Z})$, analogously to eq. \eqref{Heckedecomp}, and universality is restored.

Which are the conditions required for this to occur? In other words, for which choices of $\varOmega_{\rm SB}$ are the $\varPhi  [{g_{\rm S}\atop g_{\rm SB}}]$ constant? In order to answer this question, let us recall that a generic orbifold element $g$ involves a separate action on the left and on the right moving degrees of freedom, and thus is decomposable as $g = \gamma^{\rm L} \otimes \gamma^{\rm R}$.
In particular, $\gamma^{\rm L}$ involves the action on the RNS sector and  is the only potential source for supersymmetry breaking. Of course, $\gamma^{\rm R}$ is not arbitrary but is correlated to $\gamma^{\rm L}$ by modular invariance of the one-loop partition function, and typically involves an action on the gauge degrees of freedom. We shall assume that the orbifold acts left-right symmetrically on the six compact coordinates. As a result, $\gamma^{\rm L}$ determines the $L [{h\atop g}]$'s and, most importantly, $\gamma^{\rm R}$ determines the $\varPhi [{h\atop g}]$'s. This simple observation allows one to find the necessary conditions for universality to hold. Denote by  $g_{\rm SB} = \gamma^{\rm L}_{\rm SB} \otimes \gamma^{\rm R}$ the decomposition of the supersymmetry breaking generators of $\varOmega_{\rm SB}$. If one can replace $\gamma^{\rm L}_{\rm SB}$ with a new supersymmetry preserving action $\tilde \gamma^{\rm L}_{\rm S}$, while keeping the same action $\gamma^{\rm R}$ on the right-movers
\begin{equation}
g_{\rm SB} = \gamma^{\rm L}_{\rm SB} \otimes \gamma^{\rm R} \to \tilde g_{\rm S} = \tilde \gamma^{\rm L}_{\rm S} \otimes \gamma^{\rm R}\,,
\end{equation}
in a way that leads to a consistent supersymmetric string model, then the $\varPhi [{g_{\rm S}\atop g_{\rm SB}}]$'s are again decomposed as in \eqref{Heckedecomp}. Therefore, in the absence of extra charged massless states in the supersymmetric ``relative'' model twisted by $\varOmega_{\rm S} \times \tilde \varOmega_{\rm S}$, the $b_\mathfrak{a}$ coefficients vanish and one indeed recovers universality. 

We can summarise the above in the

\bigskip
\noindent
{\bf Universality Theorem:} {\em Any non-supersymmetric heterotic orbifold $T^6 /\varOmega_{\rm S} \times \varOmega_{\rm SB}$ yields a universal behaviour in the difference of gauge thresholds $\varDelta_{\alpha\beta}$ for gauge groups $G_\alpha$ and $G_\beta$, of rank larger than one, if $\varOmega_{\rm SB}$ can be consistently replaced by a supersymmetric orbifold $\tilde\varOmega_{\rm S}$ with the very same action on the right-moving degrees of freedom, and provided no extra  massless states charged with respect to  $G_\alpha \times G_\beta$ emerge in the bulk of the moduli space of the supersymmetric orbifold $T^6 /\varOmega_{\rm S} \times \tilde\varOmega_{\rm S}$.}
\bigskip

Which are the allowed possibilities for $\varOmega_{\rm SB}$? Recall that in symmetric orbifold constructions, fixing the action on the right moving sector automatically determines the action on the left movers as well. Therefore, the latter is either compatible with supersymmetry or not, and there is no possibility to find a $\tilde\gamma^{\rm L}_{\rm S}$ that turns $\varOmega_{\rm SB}$ to $\tilde\varOmega_{\rm S}$. This is true unless the orbifold acts trivially on the right moving $T^6$ coordinates. In this case, one has two choices which are compatible with $\gamma^{\rm R}$:  $\gamma^{\rm L} = \mathbb{1}$ that clearly preservers all supersymmetries or $\gamma^{\rm L} = (-1)^{F_{\rm st}}$, with $F_{\rm st}$ the space-time fermion number, that breaks all supersymmetries. As a result, the only way to break supersymmetry in a way which may be compatible with universality is
\begin{equation}
\varOmega_{\rm SB} = (-1)^{F_{\rm st}} \, \delta\, \gamma_{\rm gauge}\,.
\end{equation}
Here $\gamma_{\rm gauge}$ encodes the action on the Kac-Moody currents, and we have introduced an order-two shift $\delta$ along a $T^2$ to render the breaking of supersymmetry spontaneous. Without loss of generality, we shall assume that $\delta$ is a momentum shift along the $a$-cycle of the $T^2$. All the other possibilities can be reached from this one by $T$-duality transformations. $\gamma_{\rm gauge}$ is a $\mathbb{Z}_2$ generator\footnote{Because $(-1)^{F_{\rm st}}$ is order-two we choose the whole $\varOmega_{\rm SB} \sim \mathbb{Z}_2$.} and is constrained by the modular-invariance requirement \cite{Dixon:1986jc}
\begin{equation}
\sum_{I=1}^8 \left( v_I^2 + w_I^2 \right) \in \mathbb{Z}\,, \qquad {\rm with}\quad v_I , w_I \in \tfrac{1}{2} \mathbb{Z}\,.
\end{equation}
$(v_I ; w_I)$ are the eigenvalues of the orbifold acting on the complex fermions realising the ${\rm E}_8 \times {\rm E}_8$ Kac-Moody currents at the factorised point. The inequivalent solutions for $\gamma_{\rm gauge}$ are
\begin{equation}
 (0^8 ; 0^8 )_{\rm I}\,,
\qquad
(1,0^7 ; 0^8 )_{\rm II}\,,
\qquad
(1,0^7;1,0^7 )_{\rm III}\,,
\qquad
(\tfrac{1}{2}^2 ,0^6 ; \tfrac{1}{2}^2 , 0^6 )_{\rm IV}\,.
\end{equation}
The heterotic vacua $\mathcal{H}_{\rm A}$ generated by $\varOmega_{\rm SB,A} = (-1)^{f_{\rm st}}\, \delta\, \gamma_{\rm gauge, A}$, ${\rm A}={\rm I}\,,\ldots\,, {\rm IV}$, are non supersymmetric and have non-Abelian gauge groups
\begin{equation}
\begin{split}
G_{\rm I} &= {\rm E}_8 \times {\rm E}_8 \,,
\\
G_{\rm III} &= {\rm SO} (16) \times {\rm SO} (16) \,, 
\end{split}
\qquad
\begin{split}
G_{\rm II} &={\rm SO} (16)^2 \times {\rm E}_8\,,
\\
G_{\rm IV} &={\rm SU} (2) \times {\rm E}_7 \times  {\rm SU} (2) \times {\rm E}_7 \,.
\end{split}
\end{equation}
Aside from $\mathcal{H}_{\rm III}$, all these models exhibit tachyonic instabilities in some regions of the $(T,U)$ moduli space. Of course, they are all connected to one another by turning on continuous Wilson lines \cite{Nair:1986zn,Ginsparg:1986wr} but, as announced, we shall not treat them in this paper and we shall always work in regions of moduli space where the models are classically stable. 

The next step in the theorem consists in verifying if the associated supersymmetric vacuum $T^6/\varOmega_{\rm S} \times \tilde\varOmega_{\rm S}$ has or not extra charged massless states in the bulk of the $(T,U)$ moduli space. To simplify the discussion we shall assume that $\varOmega_{\rm S}$ corresponds to an orbifold limit of K3, $\varOmega_{\rm S} = T^2 \times T^4/\mathbb{Z}_N$ with $N=2,3,4,6$ and without continuous Wilson lines. As we have already stressed on several occasions, these are the only (supersymmetric) orbifolds of interest to study the moduli dependence of the threshold corrections, and the assumption to have a hard breaking of $\cN =4 \to \cN=2$ is not a limitation since, as we shall see, the $T^6/\varOmega_{\rm S} \times \tilde\varOmega_{\rm S}$ includes sectors with spontaneous $\cN =4 \to \cN=2$ breaking as well. We refer to the generators of $\varOmega_{\rm S}$ and $\tilde \varOmega_{\rm S,A}$ as $g_{\rm S}$ and $\tilde g_{\rm S,A}$, ${\rm A}= {\rm I}\,, \ldots \,, {\rm IV}$, respectively. Extra charged massless states can only emerge in the twisted sector, and actually only in the $\tilde g_{\rm S,A}$ and/or $g_{\rm S}\, \tilde g_{\rm S,A}$ ones, since it is well known that the charged spectrum of the heterotic string  compactified on the orbifold limits of K3 does not vary in the $(T,U)$ moduli space. 

To study the $\tilde g_{\rm S,A}$ twisted sector it is enough to consider the $\tilde\varOmega_{\rm S,A}$ orbifolds, since the  projection onto $g_{\rm S}$ and $g_{\rm S}\, \tilde g_{\rm S,A}$ invariant states will not generate extra massless states. In this case it is a simple exercise to show that only the orbifold $\tilde \varOmega_{\rm S,II}$ admits the enhancement of the gauge group ${\rm U }(1)^4 \times {\rm SO} (16) \times {\rm E}_8 \to {\rm U }(1)^3 \times {\rm SO} (18) \times {\rm E}_8$ at the point $U = - \bar T/2$ with $p_{\rm L} =0$. The Abelian factors are associated to the internal components of the metric and of the $B$ field with one leg along the $T^2$. In fact, the relevant contributions to the partition functions associated to the $\tilde\varOmega_{\rm S,A}$ vacua are
\begin{equation}
\begin{split}
\cZ_{\rm I} &= (V_8 - S_8 ) (\bar O_{16} + \bar S_{16} )(\bar O_{16} + \bar S_{16} ) \, \sum_{h,g=0,1} \tfrac{1}{2}\, \varGamma_{2,2} \big[{\textstyle{h\atop g}}\bigr]\,,
\\
\cZ_{\rm II} &= (V_8 - S_8 ) (\bar O_{16} + \bar S_{16} ) \bar V_{16} \, \frac{\varGamma_{2,2} [{1\atop 0}] - \varGamma_{2,2} [{1\atop 1}]}{2} + \ldots \,,
\\
\cZ_{\rm III} &= (V_8 - S_8) (\bar V_{16} \bar C_{16} + \bar C_{16} \bar V_{16} )  \, \frac{\varGamma_{2,2} [{1\atop 0}] - \varGamma_{2,2} [{1\atop 1}]}{2} + \ldots \,,
\\
\cZ_{\rm IV} &= (V_8 - S_8) \left[ ( \bar C_4 \bar O_{12} + \bar V_4 \bar S_{12} ) ( \bar S_4 \bar V_{12} + \bar O_4 \bar C_{12} ) \right.
\\
&\qquad\qquad \qquad \left.+ ( \bar S_4 \bar V_{12} + \bar O_4 \bar C_{12} )  ( \bar C_4 \bar O_{12} + \bar V_4 \bar S_{12} ) \right] \, \frac{\varGamma_{2,2} [{1\atop 0}] - \varGamma_{2,2} [{1\atop 1}]}{2} + \ldots \,.
\end{split}
\end{equation}
$\cZ_{\rm I}$ corresponds to the standard ${\rm E}_8\times {\rm E}_8$ string compactified on a $(2,2)$ lattice with new moduli $(\hat T , \hat U ) = (T/2,2U)$, while the contributions of the gauge degrees of freedom to $\cZ_{\rm III}$ and $\cZ_{\rm IV}$ are already massive and the presence of the lattice with $\hat n^1 = n^1 + \frac{1}{2}$ and odd momentum $2m_1 +1$ makes them even heavier. Only in $\cZ_{\rm II}$ the combination $\bar O_{16} \bar V_{16}$ has conformal weight $\frac{1}{2} <1$ and Kaluza Klein and winding states with $(m_1 , n^1) = (0,-1)$ and $(m_1 , n^1) = (-1,0)$ yield two extra massless vector multiplets in the fundamental representation which, together with a suitable combination of the Abelian groups, are responsible for the enhancement ${\rm U} (1) \times {\rm SO} (16) \to {\rm SO} (18)$ at the special point $U = -\bar T /2$.

The analysis of the $g_{\rm S}\, \tilde g_{\rm S,A} $ twisted sector is similar. Only for $\varOmega_{\rm S}$ orbifolds with $\mathbb{Z}_2$ sub-sectors can extra massless states emerge in the $g_{\rm S}^{[N/2]}$ twisted sector\footnote{Here we use the following definition for the symbol $[x]$: it equals $x$ if $x$ is an integer while it is zero if $x \not\in \mathbb{Z}$.}. For this reason, it suffices to consider the orbifold generated by $g_{\rm S}\, \tilde g_{\rm S,A} $, with $g_{\rm S}^2 =1$. The relevant contributions to the partition functions are
\begin{equation}
\begin{split}
\cZ_{\rm I} &= (O_4 S_4 - C_4 O_4) \, \bar O_{12} \bar S_{4} \bar O_{16}\, \frac{\varGamma_{2,2} \big[{1\atop 0}\big] - \varGamma_{2,2} \big[{1\atop 1}\big]}{2} + \ldots \,,
\\
\cZ_{\rm III} &=  (O_4 C_4 - S_4 O_4 )\, \bar O_{12} \bar S_4 \bar V_{16} \, \frac{\varGamma_{2,2} \big[{1\atop 0}\big] + \varGamma_{2,2} \big[{1\atop 1}\big]}{2} + \ldots \,.
\end{split}
\end{equation}
Clearly, model I displays the emergence of a massless hypermultiplet charged with respect to the ${\rm U} (1)$ (or ${\rm SU} (2) $ in the case $\varOmega_{\rm S} = \mathbb{Z}_2$) gauge group at the point $U=-\bar T/2$, while the spectrum of model III remains unchanged in the $(T,U)$ moduli space.  Model IV requires special care since the result depends on the way $g_{\rm S}$ and $\tilde g_{\rm S,IV}$ act on the gauge degrees of freedom. If they act on the same Cartan's of the first ${\rm E}_8$ factor, the resulting vacuum is equivalent to model I, and thus a new  charged hypermultiplet is present at the $U = -\bar T/2$ point. If  $g_{\rm S}$ and $\tilde g_{\rm S,IV}$ act along different Cartan's then no extra massless states emerge in the bulk of the $(T,U)$ moduli space. Therefore we can formulate the following 

\bigskip
\noindent
{\bf Corollary:} {\em The non-supersymmetric heterotic orbifolds $T^6 /\varOmega_{\rm S} \times \varOmega_{\rm SB}$ that yield a universal behaviour in the difference of gauge thresholds $\varDelta_{\alpha \beta}$, for gauge groups $G_\alpha$ and $G_\beta$ of rank greater than one, are $\varOmega_{\rm SB,I}$, $\varOmega_{\rm SB,III}$ and $\varOmega_{\rm SB,IV}$. In these cases,
\begin{equation}
\begin{split}
\varDelta_{\alpha \beta} &= \sum_{i=1,2,3} a_i \, \log \left[ T_2^{(i)} U_2^{(i)}\, |\eta (T^{(i)}) \eta (U^{(i)}) |^4 \right] + b_i\, \log \left[ T_2^{(i)} U_2^{(i)} \, |\vartheta_4 (T^{(i)}) \vartheta_2 (U^{(i)}) |^4 \right] 
\\
&\qquad \qquad+ c_i\, \log |j_2 (T^{(i)}/2) - j_2 (U^{(i)}) |^4\,.
\end{split} \label{coroth}
\end{equation}
The constants $a_i , b_i$ and $c_i$ are model dependent and can be computed by the knowledge of the massless spectrum only.}
\bigskip

The kind of theta constants appearing in the threshold differences depend on the choice of shift, and in \eqref{coroth} we assume a momentum shift along the $a$-cycle of the $T^2$. Different choices would have resulted in a modified argument of the theta constant, as dictated by the T-duality transformations relating the various shifts.

For instance, in case $\varOmega_{\rm S} = \mathbb{Z}_N$ corresponding to an orbifold realisation of K3 with a single $T^2$ determining the moduli dependence of the thresholds, one finds
\begin{equation}
a = \frac{\beta_\alpha^{\cN=2} - \beta_\beta^{\cN =2}}{2}\,,
\qquad
b = \beta_\alpha - \beta_\beta - \frac{\beta_\alpha^{\cN=2} - \beta_\beta^{\cN =2}}{2}\,,
\qquad
c = \delta\beta_\alpha - \delta\beta_\beta\,.
\end{equation}
Here, $\beta_\alpha^{\cN =2}$ is the beta function coefficient of the massless states of the $\cN=2$ theory generated by the $\varOmega_{\rm S}$ orbifold only, $\beta_\alpha$ is the beta function coefficient of the massless states of the full non-supersymmetric vacuum, and $\delta\beta_\alpha$ is the contribution of the extra charged massless states that may emerge at the point $p_{\rm R} =0$. The latter are associated to a non-vanishing coefficient of the $\hat\jmath_2 (\tau)$ function in the  decomposition of $L\big[{h \atop g}\big]$. These numbers can be straightforwardly extracted from \eqref{thresholds} by taking the low-energy limit, and read 
\begin{equation}
\beta_\alpha = {\rm Str} \left( \tfrac{1}{12} - s^2 \right) \, Q_\alpha^2 \,,
\end{equation}
where the supertrace is over massless states with left-moving helicity $s$ and in a given representation of the gauge group $G_\alpha$. 

Had one used a different free action of the supersymmetric breaking generator, with a different choice of shift along the $T^2$, this would have resulted in modified moduli dependence of the last two terms in \eqref{coroth}. As we shall see, this is the case in the non-supersymmetric vacua of section \ref{4210sec}.

\subsection{The special case of rank one gauge groups.}

Universality may be lost whenever one considers rank-one gauge groups. In fact, as discussed in section \ref{dissect}, the twisted vacua $|n\rangle$ of K3 orbifolds with standard embedding are necessarily charged with respect to the ${\rm U} (1)$ or ${\rm SU} (2)$  Kac-Moody groups, while being neutral with respect to the higher rank group factors.
This may have drastic consequences in the holomorphy of the threshold integrands. Of course, whenever supersymmetry is intact, these twisted vacua do not contribute to the thresholds because they are not invariant under the orbifold action and, most importantly, the left-moving sector drops out due to BPS saturation. Technically, this is a consequence of the fact that, although $\varPhi [{g_{\rm S}\atop \tilde g_{\rm S}}]$ and $\varPhi [{g_{\rm S}\atop g_{\rm S}\tilde g_{\rm S}}]$ are not constants, their sum is. However, when supersymmetry is spontaneously broken, the loss of BPS property implies that all string excitations of the left-moving sector do contribute to the thresholds, and the twisted vacua $|n\rangle$ are no longer projected away by the orbifold. As a result, the integrand in $\varDelta_{\alpha\beta}$, with one of the two groups being of rank one, involves non-holomorphic terms of the form
\begin{equation}
L \big[{\textstyle{g_{\rm S}\atop \tilde g_{\rm S}}}\big]\, \bar \varPhi \big[{\textstyle{g_{\rm S}\atop \tilde g_{\rm S}}}\big] +
L \big[{\textstyle{g_{\rm S}\atop g_{\rm S}\tilde g_{\rm S}}}\big]\, \bar \varPhi \big[{\textstyle{g_{\rm S}\atop g_{\rm S}\tilde g_{\rm S}}}\big]
\end{equation}
which destroy the holomorphy and, therefore, the universality structure.

\section{Explicit constructions}

After this general discussion, let us see how the universality theorem and its corollary work in practice in some explicit constructions. We shall consider various cases with different patterns of spontaneous (partial) supersymmetry breaking in four dimensions. In the following, in order to make the (partial) breaking of supersymmetry spontaneous we shall always employ an order-two momentum shift $\delta$ along the horizontal $a$-cycle of the $T^2$, unless otherwise stated.

\subsection{Vacua with $\cN=4 \to \cN=2$ partial supersymmetry breaking with extra massless states.}

Before we move to heterotic vacua with fully broken supersymmetry, let us first treat a case where the standard notion of universality is modified even in the presence of unbroken $\cN=2$. For concreteness, we shall focus our attention on the $T^6 /\mathbb{Z}_2 \times \mathbb{Z}_2$ orbifold, where the first factor is generated by $f$ and acts as a K3 with standard embedding, while the second one is generated by $s=(-1)^{F_1}\, \delta$, $F_1$ being the spinor number of the  ${\rm E}_8$ which is acted upon by the K3. The partition function can be written explicitly as
\begin{equation}
\begin{split}
\cZ &= \tfrac{1}{4} \sum_{h,g,H,G =0,1} \cZ \big[{\textstyle{f^{h} s^{H} \atop f^{g} s^{G}}}\big]
\\
&= \frac{1}{2^5\, \eta^{12} \bar \eta^{24}} \sum_{h,g=0,1}\sum_{H,G=0,1} \left[\sum_{a,b=0,1} (-1)^{a+b+ab} \theta \big[ {\textstyle{a/2\atop b/2}}\big]^2\theta \big[ {\textstyle{a/2+h/2\atop b/2+g/2}}\big] \theta \big[ {\textstyle{a/2-h/2\atop b/2-g/2}}\big]\right]\,
\\
&\qquad \times \left[ \sum_{k,l=0,1}   (-1)^{H l + Gk + HG}  \bar\theta \big[ {\textstyle{k/2\atop l/2}}\big]^6 \bar \theta \big[ {\textstyle{k/2+h/2\atop l/2+g/2}}\big] \bar\theta \big[ {\textstyle{k/2-h/2\atop l/2-g/2}}\big]\right]\, \left[ \sum_{c , d =0,1} \bar \theta \big[ {\textstyle{c/2\atop  d/2}}\big]^8\right]
\\
&\qquad \times \varGamma_{4,4} \big[{\textstyle{h\atop g}}\big]\, \varGamma_{2,2} \big[{\textstyle{H \atop G}}\big]\,,
\end{split}
\end{equation}
where 
\begin{equation}
\varGamma_{4,4}\big [{\textstyle{h\atop g}}\big] = \left\{ 
\begin{array}{l l}
\tau_2^{-2}\, \varGamma_{4,4} (G,B)  & {\rm for}\ (h,g)=(0,0)\,,
\\
{\displaystyle \left| \frac{4\, \eta^6 }{\theta [{1/2+h/2\atop 1/2 + g /2}]\theta [{1/2-h/2\atop 1/2 - g /2}] }\right|^2} & {\rm for}\ (h,g)\not= (0,0)\,,
\end{array}\right.
\end{equation}
is the twisted K3 lattice. The charged massless spectrum comprises $\cN=2$ vector multiplets in the adjoint representation of ${\rm U} (1)^4 \times {\rm SO} (12) \times {\rm SO} (4) \times {\rm E}_8$ and hypermultiplets in the $(\mathbold{12},\mathbold{4},1)$ representation from the untwisted sector, together with eight hypermultiplets in the $(\mathbold{12},\mathbold{2},1)$ representation from the $f$-twisted sectors. The $s$ and $sf$ twisted sectors are typically massive aside from the states
\begin{equation}
\left[ (V_4 O_4 - S_4 S_4 ) \bar V_{12} \, \bar O_4  + (O_4 V_4 - C_4 C_4 ) \bar O_{12} \, \bar V_4  + 16 (O_4 S_4 - C_4 O_4 ) \bar O_{12} \, \bar S_4\right] \, \bar O_{16}
\, \frac{\varGamma [{1\atop 0} ] - \varGamma [{1\atop 1} ]}{2}
\end{equation}
which, at the point $U = - \bar T /2$ with $p_{\rm L} =0$, contribute with extra charged vector and hyper multiplets to enhance the gauge group to 
\begin{equation}
{\rm U} (1)^4 \times {\rm SO} (12) \times {\rm SO} (4) \times {\rm E}_8 \to {\rm U} (1)^3 \times {\rm SO} (14) \times {\rm SO} (4) \times {\rm E}_8\,.
\end{equation}
This enhancement is possible due to the presence of the non-trivial discrete Wilson line induced by the $(-1)^{F_1}\delta$ orbifold and is normally forbidden in conventional constructions with trivial Wilson lines \cite{Kiritsis:1996dn}. As discussed, the presence of these extra massless states is reflected in the fact that $\varPhi$ is no longer a constant and, in particular,
$\varPhi \big[ {{f\atop s}}\big] + \varPhi \big[{{f \atop sf}}\big] $ admits the decomposition \eqref{Heckedecomp} with $b_0 = \frac{1}{2}$. 
The threshold difference for the  ${\rm E}_8$ and ${\rm SO} (12)$  gauge groups then reads
\begin{equation}
\begin{split}
\varDelta_{{\rm E}_8} - \varDelta_{{\rm SO} (12)} &= 72\, \log \left[T_2 U_2 \, |\eta (T) \eta (U) |^4 \right] + 32\, \log  \left[T_2 U_2 \, |\vartheta_4 (T) \vartheta_2 (U) |^4 \right]
\\
&\qquad -2 \, \log |j_2 (-\bar T/2 ) - j_2 (U) |^4
\end{split}
\end{equation}
and displays the emergence of the expected logarithmic singularity at the point $U = -\bar T/2$. This result extends the standard notion of universality to the case where extra charged  massless states also contribute to threshold differences.

\subsection{Vacua with $\cN =2 \to \cN =0$ spontaneous supersymmetry breaking and spectral flow} \label{20model}

The simplest prototype vacua in which non-supersymmetric universality for gauge thresholds was observed \cite{AFT} involve a spontaneously broken $\cN =2 \to \cN=0$. These are based on the $T^2 \times T^4 / \mathbb{Z}_N \times \mathbb{Z}_2$ orbifolds, where the $\mathbb{Z}_N$ is generated by $f$ and realises the K3 surface at the singular point, while the $\mathbb{Z}_2$ is generated by $s$ and is responsible for supersymmetry breaking. It can be chosen among the three cases listed in the Corollary. From the low-energy view-point these constructions correspond to flat gaugings of $\cN=2$ supergravity, thus realising the Scherk-Schwarz mechanism \cite{Scherk:1978ta,Scherk:1979zr}. This analysis is instrumental to discuss thresholds in $\cN=1 \to \cN=0$ vacua. 

For simplicity we shall restrict our analysis to the cases $\varOmega_{\rm SB,I}$ and $\varOmega_{\rm SB,III}$, since $\varOmega_{\rm SB,IV}$ is either equivalent to model I or is a straightforward generalisation. The partition function for these models reads
\begin{equation}
\begin{split}
\mathcal{Z}_F=&\frac{1}{2^4\, N\, \eta^{12} \bar\eta^{24}} \sum_{H,G=0}^1 \sum_{h,g=0}^{N-1}\left[\sum_{a,b=0}^1 (-)^{a+b+ab}
\theta \big[{\textstyle{a/2 \atop b/2}}\big]^2 \,
\theta \big[{\textstyle{a/2 +h/2 \atop b/2+g/2}}\big]\,
\theta \big[{\textstyle{a/2-h/2 \atop b/2-g/2}}\big]
\right]
\\
&\qquad\times \left[\sum_{k,l=0}^1
\bar\theta \big[{\textstyle{k/2 \atop l/2}}\big]^6\,
\bar\theta \big[{\textstyle{k/2 +h/2 \atop l/2+g/2}}\big]\,
\bar\theta \big[{\textstyle{k/2 -h/2 \atop l/2-g/2}}\big]
\right]\,
\left[\sum_{c,d=0}^1\bar\theta \big[{\textstyle{c/2 \atop d/2}}\big]^8\right] 
\\
&\qquad\times
\,(-)^{H(b+F (l+ d))+G(a+F (k+ c))+HG}\, \varGamma_{2,2} \big[{\textstyle{H\atop G}}\big]
\,\varGamma_{4,4} \big[{\textstyle{h\atop g}}\big] \,,
\end{split}\label{partition20}
\end{equation}
where $F=0,1$ corresponds to $\varOmega_{\rm SB,I}$ and $\varOmega_{\rm SB,III}$, respectively.

The threshold differences have already been computed in \cite{AFT} and are of the universal type given in eq. \eqref{coroth}, with contributions from a single $T^2$. The coefficients $a$, $b$ and $c$ can be found in \cite{AFT}. Rather than repeating the analysis of \cite{AFT}, we shall unveil the microscopic origin of the universality structure which reflects the presence of a hidden spectral flow in the right-moving sector of the heterotic string. It manifests itself in the form of a remarkable identity for the difference of non-Abelian group traces valid for any orbifold realisation of K3 and independently of the $h$ and $g$ sectors. The group traces can be conveniently written as
\begin{equation}
\varPhi_F \big[ {\textstyle{f^h s^H \atop f^g s^G}}\big] =  \frac{\varPsi_F  \big[ {\textstyle{f^h s^H \atop f^g s^G}}\big]}{\eta^{18} \theta \big[ {\textstyle{1/2+h/2\atop 1/2+g/2}} \big] \theta \big[ {\textstyle{1/2-h/2\atop 1/2-g/2}} \big] }\,,
\label{yiannis}
\end{equation}
where we have introduced for convenience
\begin{equation}
\begin{split}
\varPsi_F  \big[ {\textstyle{f^h s^H \atop f^g s^G}}\big]&=  \frac{1}{(2\pi i )^2} \,
\left( \partial^2_{x} - \partial^2_{y}\right) 
\left[ \tfrac{1}{2} \sum_{k,l=0,1} (-1)^{F (kG  + lH)}\,
\theta \big[ {\textstyle{k/2\atop l/2}}\big] (x) \theta \big[ {\textstyle{k/2\atop l/2}}\big]^5  \theta \big[ {\textstyle{k/2+h/2\atop l/2+g/2}}\big] \theta \big[ {\textstyle{k/2-h/2\atop l/2-g/2}}\big] \right]
\\
&\qquad \times \left.\left[ \tfrac{1}{2} \sum_{c,d=0,1} (-1)^{F(c G + d H)} \,\theta \big[{\textstyle{c/2 \atop d/2}}\big] (y) \theta \big[{\textstyle{c/2 \atop d/2}}\big] ^7\right]\right|_{x=0,y=0}
\\
&= \frac{i}{4 \pi}  \sum_{k,l=0,1} (-1)^{F (kG  + lH)}\,
\theta \big[ {\textstyle{k/2\atop l/2}}\big]^6  \theta \big[ {\textstyle{k/2+h/2\atop l/2+g/2}}\big] \theta \big[ {\textstyle{k/2-h/2\atop l/2-g/2}}\big] 
\\
&\qquad\times \sum_{c,d=0,1} (-1)^{F(c G + d H)} \, \theta \big[{\textstyle{c/2 \atop d/2}}\big] ^8 \, \partial_\tau \, \log \frac{\theta \big[ {\textstyle{k/2\atop l/2}}\big]}{\theta \big[{\textstyle{c/2 \atop d/2}}\big]}\,.
\end{split}
\end{equation}
In these expressions $\theta [{a\atop b}] (z)$ denotes the Jacobi theta functions with characteristics, and we have used the fact that they satisfy the Heat equation to simplify them. Although the proof that $\varPhi_F \big[ {\textstyle{f^h s^H \atop f^g s^G}}\big]$ is a constant is rather technical, it is instructive to outline the main steps since it reveals a ``hidden supersymmetry'' at work in the group trace. To this end, it is convenient to shift the dummy indices $c \to c +k $ and $d \to d +l$ and perform the sum over the new $c,d$ explicitly. This has the advantage of isolating the $\tau$ derivative, that now can be explicitly evaluated using
\begin{equation}
 \partial _\tau \log \frac{\theta\big[ {\textstyle{m/2 \atop n/2}}\big] }{\eta}= \frac{i\pi}{12}\sum_{(p,q)\not= (1,1) } (-1)^{(1+m)(1+q)} \theta \big[{\textstyle{1/2+m/2+p/2 \atop 1/2 +n/2 +q/2}}\big]^4\,,
\end{equation}
to yield
\begin{equation}
\begin{split}
\varPsi_F  \big[ {\textstyle{f^h s^H \atop f^g s^G}}\big]&=- \tfrac{1}{48} \sum_{k,l=0,1} \theta \big[{\textstyle{k/2\atop l/2}}\big]^6  \theta \big[{\textstyle{k/2+h/2\atop l/2+g/2}}\big]  \theta \big[{\textstyle{k/2-h/2\atop l/2-g/2}}\big] \Bigg[ (-1)^{FH +l} (1+ 2 (-1)^k)  \theta \big[{\textstyle{1/2\atop 0}}\big]^4  \theta \big[{\textstyle{k/2\atop 1/2+l/2}}\big]^8
\\
&\qquad - (-1)^{FG} \left( 3 (-1)^k  \theta \big[{\textstyle{0\atop 1/2}}\big]^4 + 2 (1-(-1)^l )  \theta \big[{\textstyle{1/2\atop 0}}\big]^4 \right)  \theta \big[{\textstyle{1/2+ k/2\atop l/2}}\big]^8
\\
& \qquad - (-1)^{F (H+G)} \left( 3 (-1)^k  \theta \big[{\textstyle{0\atop 0}}\big]^4 - (1+ (-1)^{k+l})  \theta \big[{\textstyle{1/2\atop 0}}\big]^4 \right)  \theta \big[{\textstyle{1/2+k/2\atop 1/2+l/2}}\big]^8 \Bigg]\,.
\end{split}
\end{equation}
One can now advocate the triple product identity
\begin{equation}
 \theta \big[{\textstyle{m/2\atop n/2}}\big]^6  \theta \big[{\textstyle{m/2+p/2\atop n/2+q/2}}\big]^6  \theta \big[{\textstyle{1/2 +p/2 \atop 1/2+q/2 }}\big]^6 = 2^6 \, \eta^{18}\,,
 \end{equation}
valid for integer values of $m,n,p,q$ provided the left-hand side does not vanish, to eliminate the common $\theta [{k/2\atop l/2}]^6$ common factor. Invoking the Jacobi identity
 \begin{equation}
 \begin{split}
& \tfrac{1}{2} \sum_{k,l=0,1} (-1)^{a l + b k} \theta \big[{\textstyle{k/2 + c/2\atop l/2 + d/2}}\big]^2 \theta \big[{\textstyle{k/2 + h/2\atop l/2 + g/2}}\big] \theta \big[{\textstyle{k/2 - h/2\atop l/2 - g/2}}\big] 
 \\
 &\qquad \qquad = (-1)^{b c} \theta \big[{\textstyle{a/2\atop b/2 + d/2}}\big]^2 \theta \big[{\textstyle{a/2 + c/2+h/2\atop b/2 + g/2}}\big] \theta \big[{\textstyle{a/2+c/2 - h/2\atop b/2 - g/2}}\big] \,,
 \end{split}\label{jacobi}
\end{equation}
which holds for generic $h,g$ and integer $a,b,c,d$, one can reorganise the expression as
\begin{equation}
\begin{split}
\varPsi_F  \big[ {\textstyle{f^h s^H \atop f^g s^G}}\big]&= - 4 \left( (-1)^{FH} + (-1)^{FG} + (-1)^{F (H+G)}\right) \eta^{18}\, \theta \big[{\textstyle{1/2 + h/2\atop 1/2 + g/2}}\big] \theta \big[{\textstyle{1/2 - h/2\atop 1/2 - g/2}}\big] 
\\
&\quad + \tfrac{8}{3}\, \eta^{18} \theta \big[{\textstyle{1/2\atop 0}}\big]^4  
\Bigg[2 \frac{(-1)^{FG}}{ \theta \big[{\textstyle{0\atop 1/2}}\big]^6} \sum_{a,b=0,1} (-1)^{a+b} \theta \big[{\textstyle{a/2\atop b/2}}\big]^2 \theta \big[{\textstyle{a/2+1/2+h/2\atop b/2+g/2}}\big] \theta \big[{\textstyle{a/2+1/2-h/2\atop b/2-g/2}}\big] 
\\
&\qquad \quad+ \frac{(-1)^{F(H+G)}}{ \theta \big[{\textstyle{0\atop 0}}\big]^6} \sum_{a,b=0,1} (-1)^{a+b} \theta \big[{\textstyle{a/2\atop b/2}}\big]^2 \theta \big[{\textstyle{a/2+1/2+h/2\atop b/2+1/2+g/2}}\big] \theta \big[{\textstyle{a/2+1/2-h/2\atop b/2 +1/2-g/2}}\big] 
\Bigg]\,.
\end{split}
\end{equation}
The first term is proportional to the denominator in \eqref{yiannis} while the terms inside the square brackets vanish as a consequence of the Jacobi identity \eqref{jacobi}. 
 This is reminiscent of how supersymmetry works in the left-moving sector of the heterotic string to yield a Fermi-Bose degenerate spectrum and indicates the presence of a similar spectral flow in the difference of gauge traces. In a sense, this is the underlying microscopic origin of universality of gauge thresholds even in the supersymmetric case. 
As a result, the contribution of the right-moving sector is constant
\begin{equation}
\varPhi_F \big[ {\textstyle{f^h s^H \atop f^g s^G}}\big] = 4- 16 \left(\frac{1+ (-1)^{FH}}{2}\right)\left(\frac{1 + (-1)^{FG}}{2} \right) \,,
\end{equation}
independently of the orbifold sectors.

\subsection{Vacua with $\cN=4 \to \cN=2 \to \cN =1 \to \cN = 0$ spontaneous supersymmetry breaking} \label{4210sec}

\begin{figure}
\centerline{
\includegraphics[width=12cm]{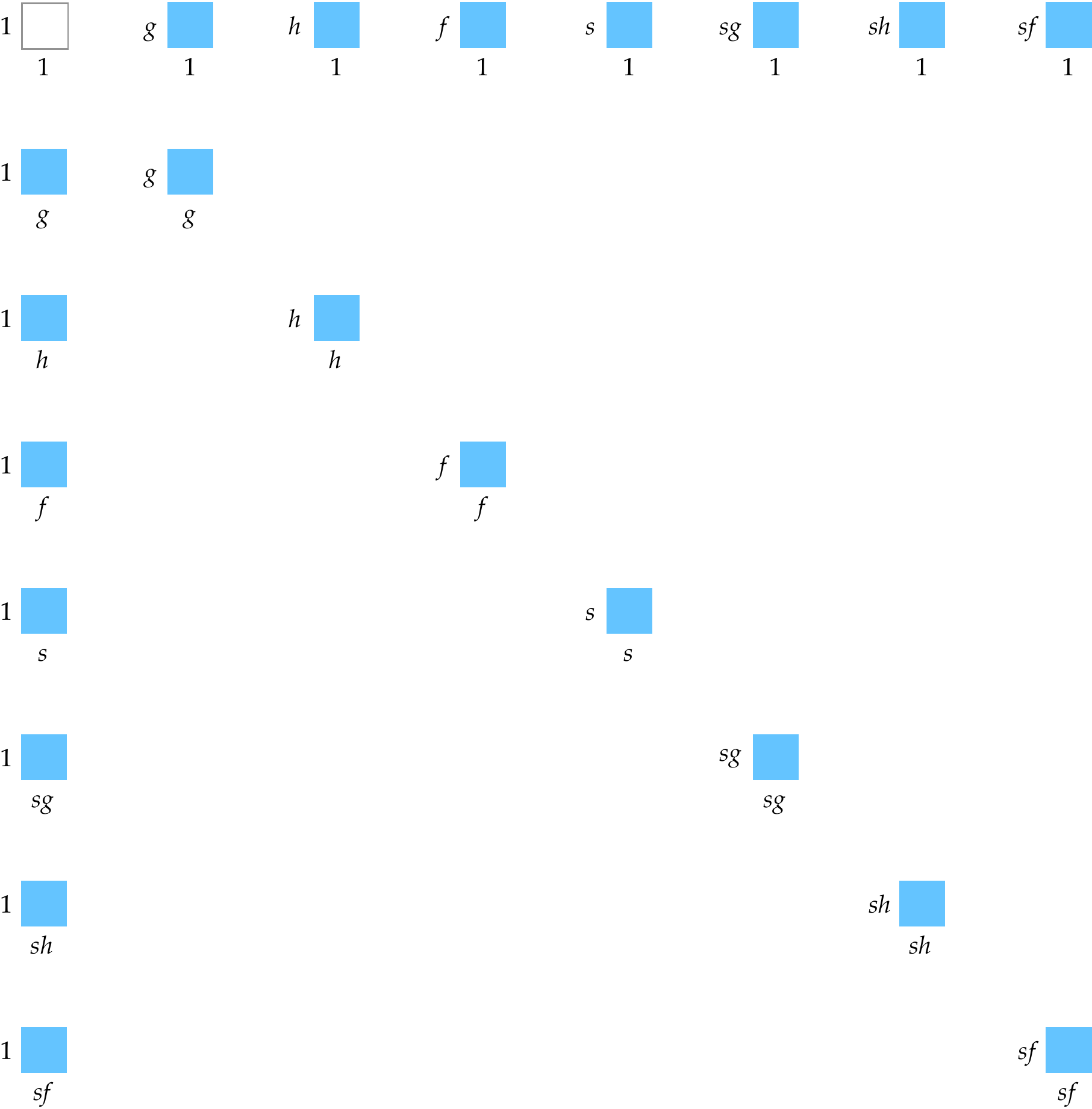}
}
\caption{The orbit structure of the $\mathbb{Z}_2 \times \mathbb{Z}_2 \times \mathbb{Z}_2$ orbifold with $\cN=4 \to \cN=2 \to \cN =1 \to \cN = 0$ spontaneously broken supersymmetry.}\label{4210model}
\end{figure}

Models with the $\cN=4 \to \cN=2 \to \cN =1$ spontaneous breaking of supersymmetry have already been studied in \cite{Kiritsis:1997ca} and can be obtained from the ${\rm E}_8 \times {\rm E}_8$ heterotic string via the action of the $\mathbb{Z}_2 \times \mathbb{Z}_2$ orbifold containing the elements
\begin{equation}
g\, : \left\{ \begin{split} z_1 &\to z_1 + \tfrac{1}{2}\,,  \\ z_2 &\to - z_2 \,, \\ z_3 &\to -z_3 \,, \end{split} \right.
\qquad
h\, : \left\{ \begin{split} z_1 &\to - z_1 \,, \\ z_2 &\to - z_2  +\tfrac{1}{2}\,, \\ z_3 &\to z_3 + \tfrac{1}{2} \,, \end{split}\right.
\qquad
f = gh \, : \left\{ \begin{split} z_1 &\to - z_1 +\tfrac{1}{2} \,, \\ z_2 &\to  z_2 + \tfrac{1}{2} \,, \\ z_3 &\to - z_3 + \tfrac{1}{2} \,. \end{split}\right.
\end{equation}
Here, $z_i$ denotes the complex coordinate on the $i^{\rm th}$ two-torus in the $T^6 = T^2 \times T^2 \times T^2$ decomposition. The presence of the shifts makes the partial breaking of supersymmetry spontaneous and, because of the simultaneous action of rotations and shifts, not all blocks in the one-loop partition function will be present. As a result, the disconnected modular orbit of this $\mathbb{Z}_2 \times \mathbb{Z}_2$ orbifold is actually absent. 

To break supersymmetry completely, we employ a further $\mathbb{Z}_2$ which is generated by $s = (-1)^{F_{\rm st} + F_1 + F_2}\, \delta_3$. Now $\delta_3$ acts as a momentum shift along the three $b$-cycles of the $T^2$'s simultaneously. A similar construction could have been obtained by using instead the generator $s = (-1)^{F_{\rm st}}\, \delta_3$. Although this would lead to a different massless spectrum, the analytic structure of the threshold differences will not be affected, and therefore we shall not discuss it explicitly. The partition function can be conveniently decomposed as
\begin{equation}
\cZ = \tfrac{1}{8} \cZ \big[{\textstyle{0\atop 0}}\big] + \tfrac{1}{4} \sum_{a}  \sum_{c,d=0,1 \atop (c,d)\not=(0,0)} \tfrac{1}{2} \cZ \big[ {\textstyle{a^c\atop a^d}} \big]
\equiv \tfrac{1}{8} \cZ \big[{\textstyle{0\atop 0}}\big] + \tfrac{1}{4} \sum_{a} \cZ_a\,,
\end{equation}
where the sum over $a$ runs over the seven non-trivial elements of the $\mathbb{Z}_2 \times \mathbb{Z}_2 \times \mathbb{Z}_2$ orbifold. The massless spectrum originates entirely from the untwisted sector, and includes the graviton, the antisymmetric Kalb-Ramond field and the dilaton as well as gauge bosons in the adjoint of ${\rm SO} (10) \times {\rm SO} (16)$, complex scalars in the representation $4 (\mathbold{1},\mathbold{1})+2 (\mathbold{10},\mathbold{1})$ and fermions in the representation $4 (\mathbold{16}, \mathbold{1}) + 4 (\mathbold{16}' , \mathbold{1}) + (\mathbold{1}, \mathbold{128})$.

Notice, that for fixed $a\not =s$, the $\cZ_a$ describe non-trivial sectors of a vacuum with $\cN =4 \to \cN=2$ spontaneously broken supersymmetry, without extra massless states in the bulk of the $(T,U)$ moduli space. Moreover, $\cZ_s$ is the only sector which is not supersymmetric but it treats  the two ${\rm E}_8$'s symmetrically so that it does not contribute to threshold differences. As a result, for the ${\rm SO} (16)$ and ${\rm SO} (10)$ gauge groups in the full vacuum with $\cN=4 \to \cN=2 \to \cN =1 \to \cN = 0$ one finds\footnote{Note that the orbits generated by $sg$, $sh$ and $sf$ involve a shift along the $a$ and $b$ cycles of the $T^2$'s. This is related to the momentum shift along the $a$ cycle by the redefinition $U \to -1/(U+1)$. At the level of the result for the thresholds, this amounts to replacing $\vartheta_2 (U)$ by $\vartheta_4 (U)$.}
\begin{equation}
\begin{split}
\varDelta_{{\rm SO} (16) - {\rm SO} (10)} &= \tfrac{1}{4} \sum_{i=1,2,3} \, \left( 48 \, \log \left[ T^{(i)}_2 U^{(i)}_2 \, |\vartheta_4 (T^{(i)}) \vartheta_2 (U^{(i)}) |^4 \right] \right.
\\
&\qquad \qquad\qquad \left. - 16 \, \log \left[ T^{(i)}_2 U^{(i)}_2 \, |\vartheta_4 (T^{(i)}) \vartheta_4 (U^{(i)}) |^4 \right]\right)\,.
\end{split}
\end{equation} 
The numerical coefficients in front of the logarithms are given by the difference of the $\beta$ functions  for the two high-rank gauge groups of the $\cN =2$ vacua $\cZ_a$, and the overall $\frac{1}{4}$ factor corresponds to the ratio of the orders of the $\mathbb{Z}_2 \times \mathbb{Z}_2 \times \mathbb{Z}_2$ and $\mathbb{Z}_2$ orbifold groups. 

\newpage

\subsection{A chiral vacuum with $\cN =1 \to \cN =0$ spontaneous supersymmetry breaking}

\begin{figure}
\centerline{
\includegraphics[width=12cm]{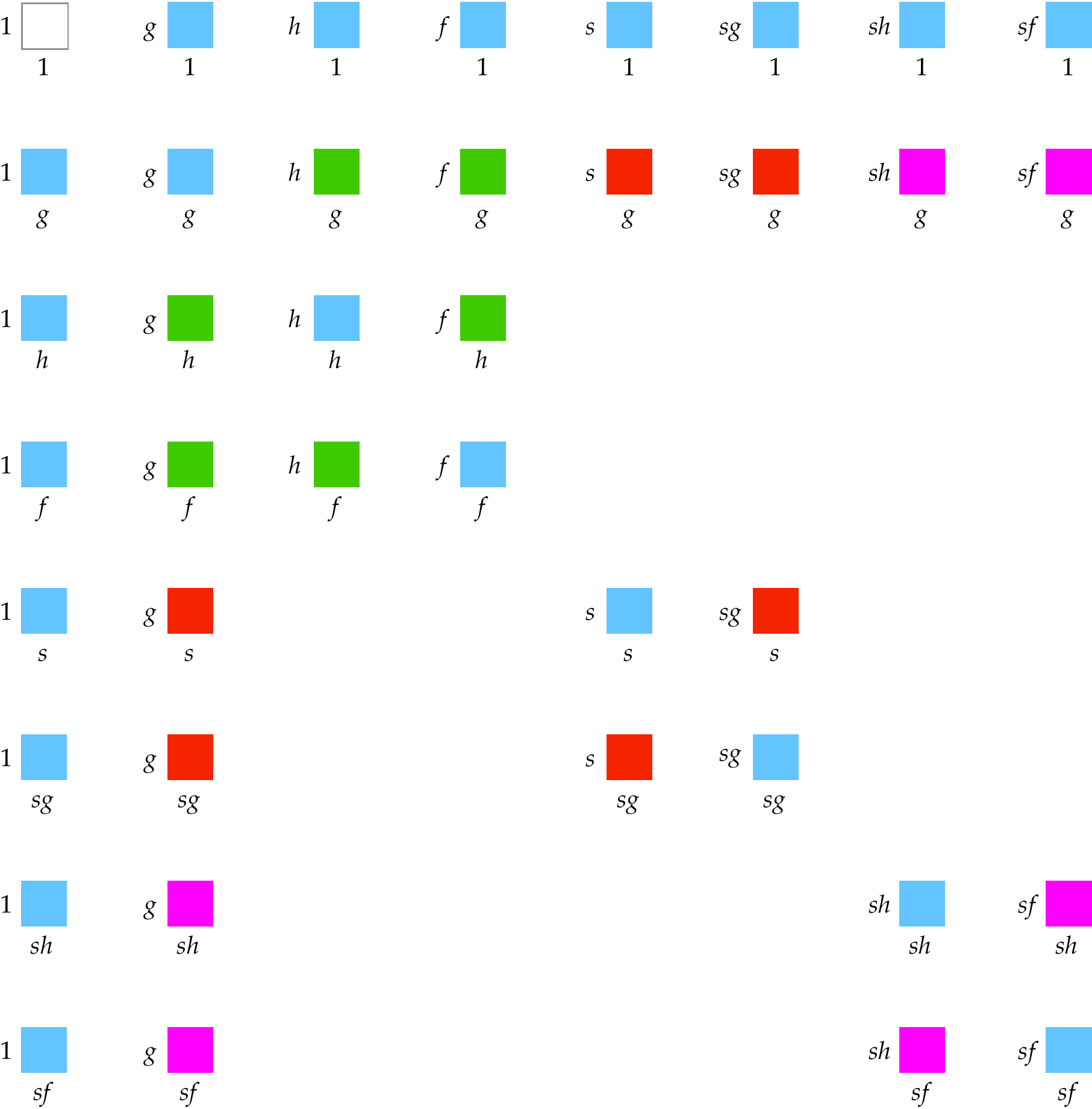}
}
\caption{The orbit structure of the $\mathbb{Z}_2 \times \mathbb{Z}_2 \times \mathbb{Z}_2$ orbifold with chiral spectrum and $\cN =1 \to \cN = 0$ spontaneously broken supersymmetry. Boxes of the same colour belong to the same modular orbit.}\label{chiral}
\end{figure}

We conclude the list of models with a four-dimensional chiral model with $\cN =1 \to \cN = 0$ spontaneously broken supersymmetry. It is again based on a $T^6/\mathbb{Z}_2 \times \mathbb{Z}_2 \times \mathbb{Z}_2$ orbifold, but now the supersymmetric $\mathbb{Z}_2 \times \mathbb{Z}_2$ action is not freely acting and is generated by
\begin{equation}
g\, : \left\{ \begin{split} z_1 &\to z_1 \,,  \\ z_2 &\to - z_2 \,, \\ z_3 &\to -z_3 \,, \end{split} \right.
\qquad
h\, : \left\{ \begin{split} z_1 &\to - z_1 \,, \\ z_2 &\to - z_2 \,, \\ z_3 &\to z_3  \,. \end{split}\right.
\end{equation}
As usual,  $f=gh$ denotes the remaining non-trivial element of the orbifold group. The spontaneous breaking $\cN =1 \to \cN =0$ is realised by the third $\mathbb{Z}_2$ factor generated by $(-1)^{F_{st}+F_1 + F_2} \delta$, with $\delta$ the usual momentum shift along the $a$-cycle of the first $T^2$, the one which is not rotated by $g$. In this case, the structure of the orbifold is richer and involves various independent modular orbits as depicted in figure \ref{chiral}.  In particular, the green orbit is responsible for the emergence of chirality in the supersymmetric $\cN=1$ (and thus $\cN=0$) vacuum. 

The non-supersymmetric massless spectrum from the untwisted sector includes the graviton, the Kalb-Ramond field, the dilaton, gauge bosons in the adjoint representation of ${\rm SO} (16) \times {\rm SO} (10 )$, one complex scalar in the representation $4 (\mathbold{1} , \mathbold{1} ) + 2 (\mathbold{1} , \mathbold{10})$ and Weyl fermions in the  representation $4 (\mathbold{1}, \mathbold{16}) + 4 (\mathbold{1}, \mathbold{16}') + (\mathbold{128}, \mathbold{1})$. The $h$ and $f$ twisted sectors have residual $\cN =1$ supersymmetry and their massless excitations comprise one complex scalar and chiral fermions in the representation $8 (\mathbold{16}, \mathbold{1}) + 8 (\mathbold{10}, \mathbold{1}) + 16 (\mathbold{1}, \mathbold{1})$. The $g$-twisted sector has broken supersymmetry and contributes with sixteen massless complex scalars in the representation $(\mathbold{10}, \mathbold{1}) + 2 (\mathbold{1}, \mathbold{1})$ together with sixteen chiral fermions in the representation $(\mathbold{16}, \mathbold{1})$. The remaining twisted sectors only contribute massive excitations.

Turning to the threshold differences for the ${\rm SO} (16)$ and ${\rm SO} (10)$ gauge groups, notice that the green and magenta orbits do not depend on the moduli of the six-torus and thus only yield a constant contribution to the thresholds, that we will not discuss. The red orbit is exactly equivalent to the red orbit in figure \ref{boxes2} and thus yields the same contribution to the threshold differences. The blue boxes involve seven independent orbits. Five correspond to standard $T^2 \times {\rm K3}$ compactifications, those associated to the $g$, $h$, $f$, $sh$ and $sf$ orbifold elements. The orbit generated by $s$ does not contribute to the threshold difference since it treats the two ${\rm E}_8$'s symmetrically. Finally, the orbit generated by $sg$ corresponds to a spontaneously broken $\cN =2 \to \cN=0$ model and thus its contribution to the thresholds is as in the previous subsections. Altogether, one gets
\begin{equation}
\begin{split}
\varDelta_{{\rm SO} (16) - {\rm SO} (10)} &= \sum_{i=1,2,3} c_i\, \log \left( T^{(i)}_2 U^{(i)}_2 \, | \eta (T^{(i)} ) \eta (U^{(i)} ) |^4\right)
\\
&\quad 
-    \tfrac{4}{3} \, \log \left( T_2^{(1)} U_2^{(1)}\, |\vartheta_4 (T^{(1)}) \vartheta_2 (U^{(1)})|^4\right) 
\\
&\quad 
+ \tfrac{1}{3} \, \log |j_2 (T^{(1)}/2 ) - j_2 (U^{(1)})|^4 \,,
\end{split}
\end{equation}
where $c_1 = 36$, $c_2 = c_3 = 24$. Again the numerical coefficients are related to the difference of $\beta$ functions as explained in previous sections.

\section{Comments on non-universal thresholds for low-rank gauge groups}

As anticipated, the threshold differences involving rank-one gauge groups are more complicated, even in vacua that meet the conditions of the universality theorem. In fact,  the twisted vacuum $|n\rangle$ in now charged with respect to this low-rank gauge groups and, unlike the supersymmetric cases, it is no longer projected away by the orbifold. This implies that the $\varPhi[{g_{\rm S}\atop g_{\rm S} g_{\rm SB}}]$ are not constants and therefore the threshold integrands are plagued by non-holomorphic contributions. Still, the latter are exponentially suppressed with the volume $T_2$ and the notion of universality does survive in the large volume limit. 

As an example, let us discuss the threshold differences for the ${\rm SO} (16)$ and ${\rm SU} (2)$ groups in the $T^6 /\mathbb{Z}_2 \times \mathbb{Z}_2$ model of \cite{AFT}, reviewed in section \ref{20model}. An explicit calculation yields 
\begin{equation}
\begin{split}
\varDelta_{{\rm SO} (16) - {\rm SU} (2)} &= -72 \int_\cF d\mu \, \varGamma_{2,2} \big[{\textstyle{0\atop 0}}\big] (T,U) + \tfrac{2}{3} \int_{\cF_2} d\mu \left( \hat\jmath (\bar\tau) -20 \right)\, \varGamma_{2,2} \big[{\textstyle{0\atop 1}}\big] (T,U) 
\\
&\qquad+\tfrac{1}{48}\int_{\cF_2} d\mu \varphi (\tau , \bar \tau ) \varGamma_{2,2} \big[{\textstyle{0\atop 1}}\big] (T,U)\,,
\end{split}
\end{equation}
with
\begin{equation}
\varphi (\tau , \bar \tau ) = \frac{\vartheta_2^8 \,(\vartheta_3^4 + \vartheta_4^4)}{\eta^{12}} \, \frac{\bar\vartheta_3^4 \, \bar\vartheta_4^4\, (\bar\vartheta_3^4 + \bar\vartheta_4^4)}{\bar\eta^{12}} = \sum_{K,L=0}^\infty c (K,L) q^{K+\frac{1}{2}} \, \bar q^{L-\frac{1}{2}}\,.
\end{equation}
The integrals in the first line can be computed straightforwardly using \cite{AFP3, AFP4} and yield the universal contributions to the threshold differences, while the non-holomorphic integral $J (T,U)$ in the second line evaluates to
\begin{equation}
\begin{split}
J (T,U) &= \sum_{m_1 , m_2 \in \mathbb{Z}} \sum_{K=0}^\infty c (K,K )\, \left( 4 (K+\tfrac{1}{2})\,  \frac{U_2}{|m_1 + \frac{1}{2} + U m_2|^2 }\right)^{1/2} \, 
\\
&\qquad\qquad\qquad \times K_1 \left( 4\pi\, \sqrt{\frac{ (K+\tfrac{1}{2}) T_2}{U_2} \, |m_1 + \tfrac{1}{2} + U m_2|^2} \right)
\\
&+\sum_{(h,g)\not=(0,0)}\sum_{0\le m_1 \le n_1 - \frac{g}{2} \atop m_2 \not=0} \sum_{K,L =0}^\infty \frac{c [{h\atop g}] (K,L)}{n_1+\frac{h}{2}}\, e^{-2\pi i (K-L+1) (m_1 +\frac{g}{2})/(n_1+\frac{h}{2})}\,
\\
&\qquad \qquad \qquad \times {\rm Li}_1 \left( e^{-2 \pi i \frac{K-L+1}{n_1 +h/2} U_1- 2\pi i (n_1 +\frac{h}{2})T_1} \, e^{-2\pi \sqrt{ (T_2 (n_1 + \frac{h}{2}))^2 +U_2^2 \frac{(K-L+1)^2}{(n_1 +h/2)^2} + 2 T_2 U_2 (K+L)} }
\right)\,,
\end{split}
\end{equation}
where $K_\nu (z)$ is the modified Bessel function of second kind, ${\rm Li}_1 (z) = - \log (1-z)$, and the coefficients $c [{0\atop 1}] (K,L) = c (K,L )$, while $c [{1\atop 0}] (K,L)$ and $c [{1\atop 1}] (K,L)$ are the Fourier coefficients of the functions obtained from $\varphi (\tau , \bar \tau) $ by an $S$ and $TS$ transformation, respectively. This expression is valid in the Weyl chamber $T_2 > U_2$, while a similar one can be obtained in the region $T_2 <U_2$.

Notice that the arguments of $K_1$ and ${\rm Li}_1$ are never vanishing due to the presence of the $\frac{1}{2}$ shift in the (Poisson summed) momenta and of the absence of the constant term in the Fourier expansion of $\varphi (\tau , \bar \tau )$. Therefore, the non universal contribution $J ( T, U)$ to the threshold differences is exponentially suppressed in the volume, and universality is actually asymptotically restored in the large $T_2$ limit even for  rank-one gauge groups.

\section{Examples without universality}

As discussed, in cases where $\varOmega_{\rm SB}$ does not meet the conditions of the universality theorem, the threshold integrands are no longer (anti-)holomorphic, and modular invariance is not sufficient to uniquely fix their form in a model independent way. However, one encounters different kinds of breaking of universality. In those cases when $\varOmega_{\rm SB}$ can be replaced by a supersymmetry preserving orbifold $\tilde\varOmega_{\rm S}$ but the latter is not protected against the emergence of extra charged massless states, a weak notion of universality is still present. In fact, in orbifold compactifications $T^6 / \varOmega_{\rm S} \times \varOmega_{\rm SB}$, one schematically finds
\begin{equation}
\begin{split}
\varDelta_{\alpha\beta} &= \varXi_{\alpha \beta} (G,B) + a\, \log \left( T_2 U_2 |\eta (T) \eta (U) |^4\right) + b \, \log \left( T_2 U_2 |\vartheta_4 (T) \vartheta_2 (U) |^4\right) 
\\
&\qquad + c \, \log | j_2 (T/2 ) - j_2 (U) |^4 + J_{\alpha\beta} (T,U)\,.
\end{split}
\end{equation}
$\varXi_{\alpha \beta}$ is non-universal contribution which depends on the $T^6$ moduli but not on the choice of $\varOmega_{\rm S}$. The term $J_{\alpha \beta} (T,U)$ is model dependent but it is exponentially suppressed in the large volume limit. As a result, this class of models is  characterised by conjugacy classes of universal behaviour depending on the choice of $\varOmega_{\rm SB}$. This is the case, for instance, for $\varOmega_{\rm SB,II}$.

Different is the situation when the supersymmetry breaking orbifold does not admit a supersymmetric completion. In this case there is no notion of universality, and the threshold differences are typically  model dependent. As an example, let us consider the orbifold $\varOmega_{\rm SB}$ generated by the element $g$ which reflects the complex coordinate of only one $T^2$  and has a similar action on the gauge degrees of freedom. Clearly, $g$ does not preserve any Killing spinor and is an order-four element when acting on the fermions. To render the breaking of supersymmetry spontaneous, we combine the action of $g$ with an order-four momentum shift along one of the remaining compact directions. The partition function of the model reads
\begin{equation}
\begin{split}
\cZ &= \frac{1}{2^5} \frac{1}{\eta^{12} \, \bar \eta^{24}} \sum_{h,g=0}^3 \left[ \sum_{a, b =0,1} (-1)^{a+b+ab} \theta \big[ {\textstyle{a/2\atop b/2}}\big]^3 \, \theta \big[ {\textstyle{a/2+h/2\atop b/2+g/2}}\big] \right]
\\
&\times \left[\sum_{k,l=0,1} \bar \theta \big[ {\textstyle{k/2\atop l/2}}\big]^7\, \bar \theta \big[ {\textstyle{k/2+h/2\atop l/2+g/2}}\big] \right] \, \left[ \sum_{c,d=0,1} \bar\theta \big[ {\textstyle{c/2\atop d/2}}\big]^8 \right]
\\
&\times e^{ i\pi kg-i \pi h (b+l)/2 }\, \varGamma_{4,4} \big[ {\textstyle{h\atop g}}\big] \, \varGamma_{2,2} \big[ {\textstyle{h\atop g}}\big]\,,
\end{split}
\end{equation}
where now it is the (4,4) lattice which is shifted, 
\begin{equation}
\varGamma_{4,4} \big[ {\textstyle{h\atop g}}\big]  = \tau_2^2 \sum_{m_i , n^i} e^{i \pi m\cdot \lambda g/2} q^{\frac{1}{4} p_{\rm L}^2} \, \bar q^{\frac{1}{4} p_{\rm R}^2}\,,
\end{equation}
with
\begin{equation}
\begin{split}
p_{{\rm L},i} &= m_i + (G +B)_{ij} (n^j+\tfrac{1}{4} h \lambda^j)\,,
\\
p_{{\rm R},i} &=m_i - (G -B)_{ij} (n^j+\tfrac{1}{4} h \lambda^j)\,,
\end{split}
\end{equation}
and $\lambda^i = (1,0,0,0)$ for a momentum shift along the first direction, while $\varGamma_{2,2} \big[ {\textstyle{h\atop g}}\big]$ is twisted,
\begin{equation}
\varGamma_{2,2}\big [{\textstyle{h\atop g}}\big] = \left\{ 
\begin{array}{l l}
\tau_2^{-1}\,\varGamma_{2,2} (g,b)  & {\rm for}\ (h,g)=(0,0)\, {\rm mod} \, 2\,,
\\
{\displaystyle \left| \frac{2\, \eta^3 }{\theta [{1/2+h/2\atop 1/2 + g /2}] }\right|^2} & {\rm for}\ (h,g)\not= (0,0) \, {\rm mod} \, 2\,.
\end{array}\right.
\end{equation}
The massless spectrum is conveniently given in terms of representations of the little group ${\rm SO} (6)$ in eight dimensions, and comprises the graviton, Kalb-Ramond and dilaton fields, together with gauge bosons in the adjoint of the ${\rm SO } (14) \times {\rm SO} (16)$ gauge group, four real scalars in the $(\mathbold{14} , \mathbold{1})$ representation, and a left-handed Weyl fermion in the $(\mathbold{64} , \mathbold{1}) + (\mathbold{64} ', \mathbold{1})$ representation. 

The threshold differences for the ${\rm SO} (16) $ and ${\rm SO} (14)$ gauge groups can be computed using eq. \eqref{thresholds}, and can be conveniently decomposed into $\varGamma_0 (4)$ and $\varGamma_0 (2)$ orbits
\begin{equation}
\varDelta_{{\rm SO} (16)-{\rm SO} (14)} \equiv \varDelta_2 + \varDelta_4\,,
\end{equation}
with
\begin{equation}
\begin{split}
\varDelta_2 &=- \tfrac{2^7}{3} \int_{\cF_2} d\mu \tau_2^{-2}\, \varGamma_{4,4} \big[ {\textstyle{0\atop 2}}\big] \varGamma_{2,2} \big[ {\textstyle{0\atop 2}}\big] \, |\xi (\tau )|^2 \,,
\\
\varDelta_4 &= -\tfrac{4}{9} \int_{\cF_4} d\mu\, \tau_2^{-1} \varGamma_{4,4} \big[ {\textstyle{0\atop 1}}\big] |\omega (\tau )|^2 \,,
\end{split}
\end{equation}
where
\begin{equation}
\begin{split}
\xi (\tau ) &= \frac{\vartheta_2^4 (\tau ) \,  \left[\vartheta_3^4 (\tau ) + \vartheta_4^4 (\tau ) \right]}{2^5 \, \eta^{12} (\tau )} \simeq 1 + 40 q + 552 q^2 + 4896 q^3 +\ldots \,,
\\
\omega (\tau ) &= \frac{\vartheta_2 ^2 (2\tau ) \, \vartheta_4^4 (2\tau ) \, \left[ 4 \vartheta_3^4 (2\tau ) + \vartheta_4^4 (2\tau ) \right]}{4\, \eta^{12} (\tau )}\simeq 
5 + 44 q + 316 q^2 + 1376 q^3 +\ldots 
\,,
\end{split}
\end{equation}
and $\cF_4$ is the fundamental domain of the Hecke congruence subgroup $\varGamma_0 (4)$. Notice that the integrands are explicitly non-holomorphic and, 
 due to  the explicit powers of $\tau_2$ in the modular integrals, the functions $\xi (\tau )$ and $\omega (\tau )$ carry now a non-trivial modular weight. This  implies that the notion of universality of eq. \eqref{Heckedecomp} is lost  and is not restored even in the large-volume limit. Indeed, standard analysis of the integrals shows that they exhibit the scaling behaviour 
 \begin{equation}
 \varDelta_2 \sim (v_2\, v_4)^{-3} \,,
 \qquad
 \varDelta_4 \sim v_4^{-2}\,,
 \end{equation}
 with $v_4$ and $v_2$ the volumes of the shifted and twisted lattices, respectively.

\section{Gravitino masses and the scales of supersymmetry breaking}

\begin{table}
\centering 
\begin{tabular}{|c | c c c c |}
\hline 
\tcolrow & & & &
\\
\tcolrow $M_o^2$ &  $\frac{1}{T_2^{(1)} U_2^{(1)}}$ & $ \frac{1}{T_2^{(2)} U_2^{(2)}} $ & $ \frac{1}{T_2^{(3)} U_2^{(3)}} $ & $\ldots $
\\[6mm]
\tcolrow $M_g^2$ &  $\frac{1}{T_2^{(1)} U_2^{(1)}}$ &  $ \frac{|1\pm U^{(2)}|^2}{T_2^{(2)} U_2^{(2)}} $ & $ \frac{|1\pm U^{(3)}|^2}{T_2^{(3)} U_2^{(3)}} $ & $\ldots $
\\[6mm]
\tcolrow $M_h^2$ &  $\frac{|1\pm U^{(1)}|^2}{T_2^{(1)} U_2^{(1)}}$ & $ \frac{|1\pm U^{(2)}|^2}{T_2^{(2)} U_2^{(2)}} $ & $ \frac{1}{T_2^{(3)} U_2^{(3)}} $ & $\ldots $
\\[6mm]
\tcolrow $M_f^2$ &  $\frac{|1\pm U^{(1)}|^2}{T_2^{(1)} U_2^{(1)}}$ & $ \frac{1}{T_2^{(2)} U_2^{(2)}} $ & $ \frac{|1\pm U^{(3)}|^2}{T_2^{(3)} U_2^{(3)}} $ & $\ldots $
\\
\tcolrow & & & &
\\
\hline
\end{tabular}
\caption{Pattern of gravitino masses for the $\cN =4 \to \cN =2 \to \cN =1 \to \cN =0$ model}    
\label{gravitino}
\end{table}

Before concluding the discussion of universality of threshold differences, let us comment on the scales of supersymmetry breaking. We shall only consider the model with $\cN=4 \to \cN =2 \to \cN =1 \to \cN =0$ breaking pattern of section \ref{4210sec}, since the others involve a single scale and their analysis is straightforward. We would like to stress that we count supersymmetries from the four-dimensional point of view.

Since any orbifold element acts freely, it is {\em a priori} impossible to identify the lightest gravitini independently of the point in moduli space. By denoting with $\psi_o$, $\psi_g$, $\psi_h$ and $\psi_f$ the four gravitini, their lightest excitations (with trivial windings) can be extracted by the following contributions to the full partition function
\begin{equation}
\begin{split}
\psi_o &\sim \tfrac{1}{4} \left[\frac{1- (-1)^{m_2 + m_4 + m_6}}{2} \, \varLambda_{m_1 , m_2}^{(1)} \varLambda_{m_ 3, m_4}^{(2)} \varLambda_{m_5 , m_6} ^{(3)} 
 + (-1)^{m_1}\, \frac{1-(-1)^{m_2}}{2}\, \varLambda_{m_1 , m_2}^{(1)} \right.
 \\
 &\quad \left.
 +  (-1)^{m_3}\, \frac{1-(-1)^{m_4}}{2}\,  \varLambda_{m_3 , m_4}^{(2)}
 +  (-1)^{m_5}\, \frac{1-(-1)^{m_6}}{2}\,  \varLambda_{m_5 , m_6}^{(3)} \right]\,,
\\
\psi_g &\sim \tfrac{1}{4} \left[\frac{1- (-1)^{m_2 + m_4 + m_6}}{2} \, \varLambda_{m_1 , m_2}^{(1)} \varLambda_{m_ 3, m_4}^{(2)} \varLambda_{m_5 , m_6} ^{(3)} 
 + (-1)^{m_1}\, \frac{1-(-1)^{m_2}}{2}\, \varLambda_{m_1 , m_2}^{(1)} \right.
 \\
 &\quad \left.
 -  (-1)^{m_3}\, \frac{1-(-1)^{m_4}}{2}\,  \varLambda_{m_3 , m_4}^{(2)}
 -  (-1)^{m_5}\, \frac{1-(-1)^{m_6}}{2}\,  \varLambda_{m_5 , m_6}^{(3)} \right]\,,
\\
\psi_h &\sim \tfrac{1}{4} \left[\frac{1- (-1)^{m_2 + m_4 + m_6}}{2} \, \varLambda_{m_1 , m_2}^{(1)} \varLambda_{m_ 3, m_4}^{(2)} \varLambda_{m_5 , m_6} ^{(3)} 
 - (-1)^{m_1}\, \frac{1-(-1)^{m_2}}{2}\, \varLambda_{m_1 , m_2}^{(1)} \right.
 \\
 &\quad \left.
 -  (-1)^{m_3}\, \frac{1-(-1)^{m_4}}{2}\,  \varLambda_{m_3 , m_4}^{(2)}
 +  (-1)^{m_5}\, \frac{1-(-1)^{m_6}}{2}\,  \varLambda_{m_5 , m_6}^{(3)} \right]\,,
\\
\psi_f &\sim \tfrac{1}{4} \left[\frac{1- (-1)^{m_2 + m_4 + m_6}}{2} \, \varLambda_{m_1 , m_2}^{(1)} \varLambda_{m_ 3, m_4}^{(2)} \varLambda_{m_5 , m_6} ^{(3)} 
 - (-1)^{m_1}\, \frac{1-(-1)^{m_2}}{2}\, \varLambda_{m_1 , m_2}^{(1)} \right.
 \\
 &\quad \left.
 +  (-1)^{m_3}\, \frac{1-(-1)^{m_4}}{2}\,  \varLambda_{m_3 , m_4}^{(2)}
 -  (-1)^{m_5}\, \frac{1-(-1)^{m_6}}{2}\,  \varLambda_{m_5 , m_6}^{(3)} \right]\,.
\end{split}
\end{equation}
In this expression we implicitly assume summation over the integers appearing in each term. Moreover, $\varLambda^{(i)}_{m_{2i-1} , m_{2i}} = \exp \{ - \pi \tau_2\, |m_{2i} - m_{2i-1} U^{(i)}|^2 \, / T_2^{(i)} U_2^{(i)} \} $. One thus gets different masses for the gravitini, and the lightest ones are summarised in table \ref{gravitino}. A close inspection of the pattern in table \ref{gravitino} shows that one can have at most three independent scales. This is consistent with the fact that the original field theoretical Scherk-Schwarz mechanism \cite{Scherk:1978ta, Scherk:1979zr} only involves three independent gravitino masses, while the  Cremmer-Scherk-Schwarz mechanism \cite{Cremmer:1979uq} with four scales is non perturbative from the string theory viewpoint. 

As expected, one recovers supersymmetry at the boundary of moduli space. However, one can only get the following patterns: $\cN =0 \to \cN = 4$ if we take, for instance, $T_2^{(1)} \to \infty$ while keeping the other moduli finite, or $\cN =0 \to \cN =2 \to \cN =4$ if we take, for instance, $U_2^{(1)} \to \infty$ first and then $T_2^{(1)} \to \infty$. 
It is thus impossible to recover minimal  $\cN=1$ supersymmetry in four dimensions. This is natural because the recovery of supersymmetry always implies the decompactification of some internal directions. As a result, the minimal number of supercharges that one can obtain depends on the number of the resulting non-compact directions, which is necessarily larger than four. This fact raises the natural question whether the Stringy Scherk-Schwarz mechanism does indeed correspond to a spontaneous breaking of supersymmetry or not. This is a well-known issue that necessitates a deeper investigation, which we do not attempt to perform here.

Before concluding, we note that in \cite{Faraggi:2014eoa} a different way to determine the scale of supersymmetry breaking/restoration was proposed. It is based on the  sector-by-sector analysis of threshold contributions and on their decoupling in the boundary of moduli space.

\section{Conclusions}

Supersymmetry breaking in String Theory is a very rich subject. Although in general the absence of supersymmetry results in total loss of control over the quantum corrections to the low-energy couplings, there exist cases which unexpectedly inherit much of the structure of their supersymmetric parents. As we have shown, this is the case for one-loop threshold differences in heterotic vacua where supersymmetry breaking is induced by a stringy Scherk-Schwarz mechanism. We have formulated the exact conditions for this to occur and shown explicitly how the very same universal structure of supersymmetric thresholds may persist also in non supersymmetric vacua. This behaviour is a consequence of a hidden symmetry acting on the right-moving (bosonic) sector of the heterotic string, and we have presented in section \ref{20model} clear evidence of its presence in the form of a generalised Jacobi identity. This extends the class of MSDS identities \cite{Kounnas:2008ft,Florakis:2009sm,Florakis:2010ty} and, as in those cases, is suggestive of the presence of a spectral flow acting on the right-moving sector.

Let us note that this is not the first example of non-supersymmetric couplings inheriting some of the non-renormalisation or stability properties of corresponding supersymmetric ones. For instance, it is known in supergravity \cite{DNP} that some four-dimensional Freund-Rubin compactifications  of eleven-dimensional supergravity admit stable non-supersym\-metric solutions which may imply that some anomalous dimensions of the holographic  (non-supersym\-metric) dual field theory living on anti-M2-branes are still protected \cite{FZ}. Another example is provided by extremal non-BPS black holes that, despite being non-supersymmetric, can still be constructed as solutions of first-order differential equations, and give rise to a notion of effective superpotential replacing the standard r\^ole of central charges \cite{CD}.

The fact that quantum corrections in non-supersymmetric theories may still share similarities with their supersymmetric parents, at least in some sectors, is remarkable and demands further investigation.

\vskip 1in 

\section*{Acknowledgements} We would like to thank Ignatios Antoniadis, Jean-Pierre Derendinger, Davide Forcella, Costas Kounnas and Herv\'e Partouche for interesting discussions. 
C.A. would like to thank the Faculty of Education Science Technology and Mathematics at the University of Canberra and the TH-Unit at CERN for hospitality during different stages of this project. I.F. would like to thank the Physics Department at the University of Torino for hospitality during different stages of this project. The work of C.A. has been partially supported by the Compagnia di San Paolo contract ``Modern Application in String Theory'' (MAST) TO-Call3-2012-0088. The work of M.T. has been partially supported by the Australian Research Council grant DP120101340. M.T. would also like to acknowledge the grant 31/89 of the Shota Rustaveli National Science Foundation.

\bibliographystyle{unsrt}

\begin{thebibliography}{99}


\bibitem{Rohm:1983aq}
  R.~Rohm,
  ``Spontaneous Supersymmetry Breaking in Supersymmetric String Theories,''
  Nucl.\ Phys.\ B {\bf 237} (1984) 553.
  
\bibitem{SSstringii}
C.~Kounnas and M.~Porrati,
``Spontaneous Supersymmetry Breaking In String Theory,''
Nucl.\ Phys.\ B {\bf 310} (1988) 355.

\bibitem{FKPZ}
S.~Ferrara, C.~Kounnas, M.~Porrati and F.~Zwirner,
``Superstrings With Spontaneously Broken Supersymmetry And Their 
Effective Theories,''
Nucl.\ Phys.\ B {\bf 318} (1989) 75.
  
  \bibitem{Kounnas:1989dk}
  C.~Kounnas and B.~Rostand,
  ``Coordinate Dependent Compactifications and Discrete Symmetries,''
  Nucl.\ Phys.\ B {\bf 341} (1990) 641.
  
   \bibitem{AlvarezGaume:1986jb}
  L.~Alvarez-Gaume, P.~H.~Ginsparg, G.~W.~Moore and C.~Vafa,
  ``An ${\rm O}(16) \times {\rm O}(16)$ Heterotic String,''
  Phys.\ Lett.\ B {\bf 171} (1986) 155.

  \bibitem{Dixon:1986iz}
  L.~J.~Dixon and J.~A.~Harvey,
  ``String Theories in Ten-Dimensions Without Space-Time Supersymmetry,''
  Nucl.\ Phys.\ B {\bf 274} (1986) 93.

    \bibitem{Ginsparg:1986wr}
  P.~H.~Ginsparg and C.~Vafa,
  ``Toroidal Compactification of Nonsupersymmetric Heterotic Strings,''
  Nucl.\ Phys.\ B {\bf 289} (1987) 414.
  
  \bibitem{Nair:1986zn}
  V.~P.~Nair, A.~D.~Shapere, A.~Strominger and F.~Wilczek,
  ``Compactification of the Twisted Heterotic String,''
  Nucl.\ Phys.\ B {\bf 287} (1987) 402.

  \bibitem{Itoyama:1986ei}
  H.~Itoyama and T.~R.~Taylor,
  ``Supersymmetry Restoration in the Compactified ${\rm O}(16) \times {\rm O}(16)'$ Heterotic String Theory,''
  Phys.\ Lett.\ B {\bf 186} (1987) 129.
  
  
  \bibitem{Taylor:1987uv}
  T.~R.~Taylor,
  ``Model Building on Asymmetric $\mathbb{Z}_3$ Orbifolds: Nonsupersymmetric Models,''
  Nucl.\ Phys.\ B {\bf 303} (1988) 543.
  
  \bibitem{Toon:1990ij}
  A.~Toon,
  ``Nonsupersymmetric $\mathbb{Z}_4$ Orbifolds And Atkin-lehner Symmetry,''
  Phys.\ Lett.\ B {\bf 243} (1990) 68.
  
   \bibitem{Dienes:1995bx}
  K.~R.~Dienes and A.~E.~Faraggi,
  ``Gauge coupling unification in realistic free fermionic string models,''
  Nucl.\ Phys.\ B {\bf 457} (1995) 409
  [hep-th/9505046].

\bibitem{Sasada:1995wq}
  T.~Sasada,
  ``Asymmetric orbifold models of nonsupersymmetric heterotic strings,''
  Prog.\ Theor.\ Phys.\  {\bf 95} (1996) 249
  [hep-th/9508098].

  \bibitem{Kiritsis:1997ca}
  E.~Kiritsis and C.~Kounnas,
  ``Perturbative and nonperturbative partial supersymmetry breaking: $\cN=4 \to \cN= 2 \to \cN=1$,''
  Nucl.\ Phys.\ B {\bf 503} (1997) 117
  [hep-th/9703059].
    
    \bibitem{Blum:1997cs}
  J.~D.~Blum and K.~R.~Dienes,
  ``Duality without supersymmetry: The Case of the ${\rm SO}(16) \times {\rm SO}(16)$ string,''
  Phys.\ Lett.\ B {\bf 414} (1997) 260
  [hep-th/9707148].

   \bibitem{Blum:1997gw}
  J.~D.~Blum and K.~R.~Dienes,
  ``Strong / weak coupling duality relations for nonsupersymmetric string theories,''
  Nucl.\ Phys.\ B {\bf 516} (1998) 83
  [hep-th/9707160].
  
   \bibitem{Harvey:1998rc}
  J.~A.~Harvey,
  ``String duality and nonsupersymmetric strings,''
  Phys.\ Rev.\ D {\bf 59} (1999) 026002
  [hep-th/9807213].

     \bibitem{Ghilencea:2001bv}
  D.~M.~Ghilencea, H.~P.~Nilles and S.~Stieberger,
  ``Divergences in Kaluza-Klein models and their string regularization,''
  New J.\ Phys.\  {\bf 4} (2002) 15
  [hep-th/0108183].
  
  \bibitem{Font:2002pq}
  A.~Font and A.~Hernandez,
  ``Nonsupersymmetric orbifolds,''
  Nucl.\ Phys.\ B {\bf 634} (2002) 51
  [hep-th/0202057].

  \bibitem{Faraggi:2007tj}
  A.~E.~Faraggi and M.~Tsulaia,
  ``On the Low Energy Spectra of the Nonsupersymmetric Heterotic String Theories,''
  Eur.\ Phys.\ J.\ C {\bf 54} (2008) 495
  [arXiv:0706.1649 [hep-th]].
 
  \bibitem{Florakis:2009sm}
  I.~Florakis and C.~Kounnas,
 ``Orbifold Symmetry Reductions of Massive Boson-Fermion Degeneracy,''
  Nucl.\ Phys.\ B {\bf 820} (2009) 237
  [arXiv:0901.3055 [hep-th]].
  
    \bibitem{Faraggi:2009xy}
  A.~E.~Faraggi and M.~Tsulaia,
  ``Interpolations Among NAHE-based Supersymmetric and Nonsupersymmetric String Vacua,''
  Phys.\ Lett.\ B {\bf 683} (2010) 314
  [arXiv:0911.5125 [hep-th]].
  
  \bibitem{Florakis:2010ty}
  I.~Florakis, C.~Kounnas and N.~Toumbas,
  ``Marginal Deformations of Vacua with Massive boson-fermion Degeneracy Symmetry,''
  Nucl.\ Phys.\ B {\bf 834} (2010) 273
  [arXiv:1002.2427 [hep-th]].
 
  
 
   
    \bibitem{AS}
  A.~Sagnotti,
  ``Surprises in open string perturbation theory,''
  Nucl.\ Phys.\ Proc.\ Suppl.\  {\bf 56B} (1997) 332
  [hep-th/9702093].

  \bibitem{Antoniadis:1998ki}
  I.~Antoniadis, E.~Dudas and A.~Sagnotti,
  ``Supersymmetry breaking, open strings and M theory,''
  Nucl.\ Phys.\ B {\bf 544} (1999) 469
  [hep-th/9807011].

  \bibitem{Angelantonj:1998gj}
  C.~Angelantonj,
  ``Nontachyonic open descendants of the 0B string theory,''
  Phys.\ Lett.\ B {\bf 444} (1998) 309
  [hep-th/9810214].

  \bibitem{Antoniadis:1998ep}
  I.~Antoniadis, G.~D'Appollonio, E.~Dudas and A.~Sagnotti,
  ``Partial breaking of supersymmetry, open strings and M theory,''
  Nucl.\ Phys.\ B {\bf 553} (1999) 133
  [hep-th/9812118].

  \bibitem{Blumenhagen:1999ns}
  R.~Blumenhagen, A.~Font and D.~Lust,
  ``Tachyon free orientifolds of type 0B strings in various dimensions,''
  Nucl.\ Phys.\ B {\bf 558} (1999) 159
  [hep-th/9904069].

  \bibitem{Angelantonj:1999gm}
  C.~Angelantonj, I.~Antoniadis and K.~Forger,
  ``Nonsupersymmetric type I strings with zero vacuum energy,''
  Nucl.\ Phys.\ B {\bf 555} (1999) 116
  [hep-th/9904092].

  \bibitem{Sugimoto:1999tx}
  S.~Sugimoto,
  ``Anomaly cancellations in type I D-9 - anti-D-9 system and the USp(32) string theory,''
  Prog.\ Theor.\ Phys.\  {\bf 102} (1999) 685
  [hep-th/9905159].

  \bibitem{Antoniadis:1999ux}
  I.~Antoniadis, G.~D'Appollonio, E.~Dudas and A.~Sagnotti,
  ``Open descendants of $\mathbb{Z}_2 \times \mathbb{Z}_2$ freely acting orbifolds,''
  Nucl.\ Phys.\ B {\bf 565} (2000) 123
  [hep-th/9907184].

  \bibitem{Antoniadis:1999xk}
  I.~Antoniadis, E.~Dudas and A.~Sagnotti,
  ``Brane supersymmetry breaking,''
  Phys.\ Lett.\ B {\bf 464} (1999) 38
  [hep-th/9908023].

  \bibitem{Angelantonj:1999jh}
  C.~Angelantonj,
  ``Comments on open string orbifolds with a nonvanishing $B_{ab}$,''
  Nucl.\ Phys.\ B {\bf 566} (2000) 126
  [hep-th/9908064].

   \bibitem{Aldazabal:1999jr}
  G.~Aldazabal and A.~M.~Uranga,
  ``Tachyon free nonsupersymmetric type IIB orientifolds via Brane - anti-brane systems,''
  JHEP {\bf 9910} (1999) 024
  [hep-th/9908072].

   \bibitem{Angelantonj:1999ms}
  C.~Angelantonj, I.~Antoniadis, G.~D'Appollonio, E.~Dudas and A.~Sagnotti,
  ``Type I vacua with brane supersymmetry breaking,''
  Nucl.\ Phys.\ B {\bf 572} (2000) 36
  [hep-th/9911081].

  \bibitem{Borunda:2002ra}
  M.~Borunda, M.~Serone and M.~Trapletti,
  ``On the quantum stability of IIB orbifolds and orientifolds with Scherk-Schwarz SUSY breaking,''
  Nucl.\ Phys.\ B {\bf 653} (2003) 85
  [hep-th/0210075].

  \bibitem{Angelantonj:2003hr}
  C.~Angelantonj and I.~Antoniadis,
  ``Suppressing the cosmological constant in nonsupersymmetric type I strings,''
  Nucl.\ Phys.\ B {\bf 676} (2004) 129
  [hep-th/0307254].

   \bibitem{Angelantonj:2005hs}
  C.~Angelantonj, M.~Cardella and N.~Irges,
  ``Scherk-Schwarz breaking and intersecting branes,''
  Nucl.\ Phys.\ B {\bf 725} (2005) 115
  [hep-th/0503179].
  
  \bibitem{Angelantonj:2006ut}
  C.~Angelantonj, M.~Cardella and N.~Irges,
  ``An Alternative for Moduli Stabilisation,''
  Phys.\ Lett.\ B {\bf 641} (2006) 474
  [hep-th/0608022].
  
  \bibitem{GatoRivera:2007yi}
  B.~Gato-Rivera and A.~N.~Schellekens,
  ``Non-supersymmetric Tachyon-free Type-II and Type-I Closed Strings from RCFT,''
  Phys.\ Lett.\ B {\bf 656} (2007) 127
  [arXiv:0709.1426 [hep-th]].
  
  \bibitem{GatoRivera:2008zn}
  B.~Gato-Rivera and A.~N.~Schellekens,
  ``Non-supersymmetric Orientifolds of Gepner Models,''
  Phys.\ Lett.\ B {\bf 671} (2009) 105
  [arXiv:0810.2267 [hep-th]].
  
  \bibitem{Blaszczyk:2014qoa}
  M.~Blaszczyk, S.~Groot Nibbelink, O.~Loukas and S.~Ramos-Sanchez,
  ``Non-supersymmetric heterotic model building,''
  JHEP {\bf 1410} (2014) 119
  [arXiv:1407.6362 [hep-th]].

\bibitem{Shiu:1998he}
  G.~Shiu and S.~H.~H.~Tye,
  ``Bose-Fermi degeneracy and duality in nonsupersymmetric strings,''
  Nucl.\ Phys.\ B {\bf 542} (1999) 45
  [hep-th/9808095].

  \bibitem{Abel:2015oxa}
  S.~Abel, K.~R.~Dienes and E.~Mavroudi,
  ``Towards a nonsupersymmetric string phenomenology,''
  Phys.\ Rev.\ D {\bf 91} (2015) 12,  126014
  [arXiv:1502.03087 [hep-th]].
  
  \bibitem{Lukas:2015kca}
  A.~Lukas, Z.~Lalak and E.~E.~Svanes,
  ``Heterotic Moduli Stabilisation and Non-Supersymmetric Vacua,''
  JHEP {\bf 1508} (2015) 020
  [arXiv:1504.06978 [hep-th]].

  \bibitem{Kutasov:1990sv}
  D.~Kutasov and N.~Seiberg,
  ``Number of degrees of freedom, density of states and tachyons in string theory and CFT,''
  Nucl.\ Phys.\ B {\bf 358} (1991) 600.
  
  \bibitem{Dienes:1994np}
  K.~R.~Dienes,
  ``Modular invariance, finiteness, and misaligned supersymmetry: New constraints on the numbers of physical string states,''
  Nucl.\ Phys.\ B {\bf 429} (1994) 533
  [hep-th/9402006].
  
  \bibitem{Angelantonj:2010ic}
  C.~Angelantonj, M.~Cardella, S.~Elitzur and E.~Rabinovici,
  ``Vacuum stability, string density of states and the Riemann zeta function,''
  JHEP {\bf 1102} (2011) 024
  [arXiv:1012.5091 [hep-th]].

\bibitem{Hagedorn:1965st}
  R.~Hagedorn,
  ``Statistical thermodynamics of strong interactions at high-energies,''
  Nuovo Cim.\ Suppl.\  {\bf 3} (1965) 147.
  
  \bibitem{Atick:1988si}
  J.~J.~Atick and E.~Witten,
  ``The Hagedorn Transition and the Number of Degrees of Freedom of String Theory,''
  Nucl.\ Phys.\ B {\bf 310} (1988) 291.
  
  \bibitem{Antoniadis:1991kh}
  I.~Antoniadis and C.~Kounnas,
  ``Superstring phase transition at high temperature,''
  Phys.\ Lett.\ B {\bf 261} (1991) 369.


  \bibitem{Ashfaque:2015vta}
  J.~M.~Ashfaque, P.~Athanasopoulos, A.~E.~Faraggi and H.~Sonmez,
  ``Non-Tachyonic Semi-Realistic Non-Supersymmetric Heterotic String Vacua,''
  arXiv:1506.03114 [hep-th].

  \bibitem{Blaszczyk:2015zta}
  M.~Blaszczyk, S.~G.~Nibbelink, O.~Loukas and F.~Ruehle,
  ``Calabi-Yau compactifications of non-supersymmetric heterotic string theory,''
  arXiv:1507.06147 [hep-th].
  
  \bibitem{Nibbelink:2015vha}
  S.~G.~Nibbelink, O.~Loukas and F.~Ruehle,
  ``(MS)SM-like models on smooth Calabi-Yau manifolds from all three heterotic string theories,''
  arXiv:1507.07559 [hep-th].
  
  \bibitem{AFT}
  C.~Angelantonj, I.~Florakis and M.~Tsulaia,
  ``Universality of Gauge Thresholds in Non-Supersymmetric Heterotic Vacua,''
  Phys.\ Lett.\ B {\bf 736} (2014) 365
  [arXiv:1407.8023 [hep-th]].

  \bibitem{Dixon:1990pc}
  L.~J.~Dixon, V.~Kaplunovsky and J.~Louis,
  ``Moduli dependence of string loop corrections to gauge coupling constants,''
  Nucl.\ Phys.\ B {\bf 355} (1991) 649.
  
  \bibitem{Mayr:1993mq}
  P.~Mayr and S.~Stieberger,
  ``Threshold corrections to gauge couplings in orbifold compactifications,''
  Nucl.\ Phys.\ B {\bf 407} (1993) 725
  [hep-th/9303017].
  
   \bibitem{Kiritsis:1996dn}
  E.~Kiritsis, C.~Kounnas, P.~M.~Petropoulos and J.~Rizos,
  ``Universality properties of $\cN=2$ and $\cN=1$ heterotic threshold corrections,''
  Nucl.\ Phys.\ B {\bf 483} (1997) 141
  [hep-th/9608034].
  
  \bibitem{Faraggi:2011aw}
  A.~E.~Faraggi, I.~Florakis, T.~Mohaupt and M.~Tsulaia,
  ``Conformal Aspects of Spinor-Vector Duality,''
  Nucl.\ Phys.\ B {\bf 848} (2011) 332
  [arXiv:1101.4194 [hep-th]].
  
  \bibitem{AFP3}
  C.~Angelantonj, I.~Florakis and B.~Pioline,
  ``Rankin-Selberg methods for closed strings on orbifolds,''
  JHEP {\bf 1307} (2013) 181
  [arXiv:1304.4271 [hep-th]].
  
  \bibitem{Dixon:1986qv}
  L.~J.~Dixon, D.~Friedan, E.~J.~Martinec and S.~H.~Shenker,
  ``The Conformal Field Theory of Orbifolds,''
  Nucl.\ Phys.\ B {\bf 282} (1987) 13.
  
  \bibitem{Hamidi:1986vh}
  S.~Hamidi and C.~Vafa,
  ``Interactions on Orbifolds,''
  Nucl.\ Phys.\ B {\bf 279} (1987) 465.
  
   \bibitem{Kaplunovsky:1987rp}
  V.~S.~Kaplunovsky,
  ``One Loop Threshold Effects in String Unification,''
  Nucl.\ Phys.\ B {\bf 307} (1988) 145
   [Nucl.\ Phys.\ B {\bf 382} (1992) 436]
  [hep-th/9205068].
 
   \bibitem{Antoniadis:1992pm}
  I.~Antoniadis, E.~Gava, K.~S.~Narain and T.~R.~Taylor,
  ``Superstring threshold corrections to Yukawa couplings,''
  Nucl.\ Phys.\ B {\bf 407} (1993) 706
  [hep-th/9212045].
 
  \bibitem{LopesCardoso:1994vn}
  G.~Lopes Cardoso, D.~Lust and T.~Mohaupt,
  ``Threshold corrections and symmetry enhancement in string compactifications,''
  Nucl.\ Phys.\ B {\bf 450} (1995) 115
  [hep-th/9412209].
  
  \bibitem{Mayr:1995rx}
  P.~Mayr and S.~Stieberger,
  ``Moduli dependence of one loop gauge couplings in (0,2) compactifications,''
  Phys.\ Lett.\ B {\bf 355} (1995) 107
  [hep-th/9504129].
  
  \bibitem{Harvey:1995fq}
  J.~A.~Harvey and G.~W.~Moore,
  ``Algebras, BPS states, and strings,''
  Nucl.\ Phys.\ B {\bf 463} (1996) 315
  [hep-th/9510182].
  
  \bibitem{LopesCardoso:1996nc}
  G.~Lopes Cardoso, G.~Curio and D.~Lust,
  ``Perturbative couplings and modular forms in $\cN=2$ string models with a Wilson line,''
  Nucl.\ Phys.\ B {\bf 491} (1997) 147
  [hep-th/9608154].
  
  \bibitem{Gregori:1997hi}
  A.~Gregori, E.~Kiritsis, C.~Kounnas, N.~A.~Obers, P.~M.~Petropoulos and B.~Pioline,
  ``$R^2$ corrections and nonperturbative dualities of $\cN=4$ string ground states,''
  Nucl.\ Phys.\ B {\bf 510} (1998) 423
  [hep-th/9708062].
  
   \bibitem{AFP1}
  C.~Angelantonj, I.~Florakis and B.~Pioline,
  ``A new look at one-loop integrals in string theory,''
  Commun.\ Num.\ Theor.\ Phys.\  {\bf 6} (2012) 159
  [arXiv:1110.5318 [hep-th]].
  
  \bibitem{AFP2}
  C.~Angelantonj, I.~Florakis and B.~Pioline,
  ``One-Loop BPS amplitudes as BPS-state sums,''
  JHEP {\bf 1206} (2012) 070
  [arXiv:1203.0566 [hep-th]].
  
   \bibitem{Kiritsis:1998en}
  E.~Kiritsis, C.~Kounnas, P.~M.~Petropoulos and J.~Rizos,
  ``String threshold corrections in models with spontaneously broken supersymmetry,''
  Nucl.\ Phys.\ B {\bf 540} (1999) 87
  [hep-th/9807067].
  
 \bibitem{AFP4}
  C.~Angelantonj, I.~Florakis and B.~Pioline,
  ``Threshold corrections, generalised prepotentials and Eichler integrals,''
  Nucl.\ Phys.\ B {\bf 897} (2015) 781
  [arXiv:1502.00007 [hep-th]].
  
  \bibitem{Dixon:1986jc}
  L.~J.~Dixon, J.~A.~Harvey, C.~Vafa and E.~Witten,
  ``Strings on Orbifolds. 2.,''
  Nucl.\ Phys.\ B {\bf 274} (1986) 285.
  
  \bibitem{Scherk:1978ta}
  J.~Scherk and J.~H.~Schwarz,
  ``Spontaneous Breaking of Supersymmetry Through Dimensional Reduction,''
  Phys.\ Lett.\ B {\bf 82} (1979) 60.
  
  \bibitem{Scherk:1979zr}
  J.~Scherk and J.~H.~Schwarz,
  ``How to Get Masses from Extra Dimensions,''
  Nucl.\ Phys.\ B {\bf 153} (1979) 61.
  
  \bibitem{Cremmer:1979uq}
  E.~Cremmer, J.~Scherk and J.~H.~Schwarz,
  ``Spontaneously Broken $\cN=8$ Supergravity,''
  Phys.\ Lett.\ B {\bf 84} (1979) 83.
  
  \bibitem{Faraggi:2014eoa}
  A.~E.~Faraggi, C.~Kounnas and H.~Partouche,
  ``Large volume susy breaking with a solution to the decompactification problem,''
  Nucl.\ Phys.\ B {\bf 899} (2015) 328
  [arXiv:1410.6147 [hep-th]].

\bibitem{Kounnas:2008ft}
  C.~Kounnas,
  ``Massive Boson-Fermion Degeneracy and the Early Structure of the Universe,''
  Fortsch.\ Phys.\  {\bf 56} (2008) 1143
  [arXiv:0808.1340 [hep-th]].
  
  \bibitem{DNP}
  M.~J.~Duff, B.~E.~W.~Nilsson and C.~N.~Pope,
  ``The Criterion for Vacuum Stability in Kaluza-Klein Supergravity,''
  Phys.\ Lett.\ B {\bf 139} (1984) 154.

\bibitem{FZ}
  D.~Forcella and A.~Zaffaroni,
  ``Non-supersymmetric CS-matter theories with known AdS duals,''
  Adv.\ High Energy Phys.\  {\bf 2011} (2011) 393645
  [arXiv:1103.0648 [hep-th]].

\bibitem{CD}
  A.~Ceresole and G.~Dall'Agata,
  ``Flow Equations for Non-BPS Extremal Black Holes,''
  JHEP {\bf 0703} (2007) 110
  [hep-th/0702088].
 \end{thebibliography}

\end{document}